\documentclass[conference]{IEEEtran}

\usepackage{amsmath}
\usepackage{fixltx2e}
\usepackage{hyperref}
\usepackage{textcomp}
\usepackage{listings}
\usepackage{array}
\usepackage[utf8]{inputenc}
\usepackage{xcolor}
\usepackage{multirow}
\usepackage[obeyspaces]{xurl}
\usepackage{graphicx}
\usepackage{xspace}
\usepackage{caption}
\usepackage{amssymb}
\usepackage{algorithm}
\usepackage{algorithmic}
\usepackage{hhline}
\usepackage{makecell}
\usepackage{amssymb}
\usepackage{bbding}

\ifCLASSINFOpdf
\else
\fi

\graphicspath{ {figures/} }

\newcommand{\sysname}{D-Box\xspace}
\newcommand{\fmpu}{F-MPU\xspace}

\newcommand{\dctr}{DMA controller\xspace}
\newcommand{\dctrs}{DMA controllers\xspace}
\newcommand{\Circled}[1]{\textcircled{\footnotesize{#1}}}

\begin{document}
\title{\sysname: DMA-enabled Compartmentalization for Embedded Applications}

\author{\IEEEauthorblockN{
Alejandro Mera, 
Yi Hui Chen, 
Ruimin Sun, 
Engin Kirda and  
Long Lu
}
\IEEEauthorblockA{Northeastern University\\
\{mera.a, chen.yihui, r.sun, e.kirda and l.lu\}@northeastern.edu}
}

\IEEEoverridecommandlockouts
\makeatletter\def\@IEEEpubidpullup{6.5\baselineskip}\makeatother
\IEEEpubid{\parbox{\columnwidth}{
    Network and Distributed Systems Security (NDSS) Symposium 2022\\
    27 February - 3 March 2022, San Diego, CA, USA\\
    ISBN 1-891562-74-6\\
    https://dx.doi.org/10.14722/ndss.2022.24053\\
    www.ndss-symposium.org
}
\hspace{\columnsep}\makebox[\columnwidth]{}}

\maketitle

\begin{abstract}
Embedded and Internet-of-Things (IoT) devices have seen an increase
in adoption in many domains. The security of these devices is of great
importance as they are often used to control critical infrastructure,
medical devices, and vehicles.  Existing solutions
to isolate microcontroller (MCU) resources in order to increase
their security face significant challenges such as specific hardware
unavailability, Memory Protection Unit (MPU) limitations
and a significant lack of Direct Memory Access (DMA) support.
Nevertheless, DMA is fundamental for the 
power and performance requirements of embedded applications.

In this paper, we present \sysname, a systematic approach to 
enable secure DMA operations for compartmentalization solutions of 
embedded applications using real-time operating systems (RTOS). 
\sysname defines a reference architecture and a workflow 
to protect DMA operations holistically. It provides practical methods 
to harden the kernel and define capability-based security policies 
for easy definition of DMA operations 
with strong security properties. We implemented a \sysname prototype
for the Cortex-M3/M4 on top of the popular FreeRTOS-MPU (\fmpu). 
The \sysname procedures and a stricter security model enabled 
DMA operations, yet it exposed 41 times less ROP (return-orienting-programming) 
gadgets when compared with the standard \fmpu. \sysname adds only a 2\%
processor overhead while reducing the power consumption of peripheral
operation benchmarks by 18.2\%. The security properties and 
performance of \sysname were tested and confirmed on 
a real-world case study of a Programmable Logic 
Controller (PLC) application. 

\end{abstract}

\section{Introduction}\label{sec:introduction}

Embedded and IoT devices are increasingly becoming popular
\cite{pelinoPredictions2021Technology2020}.  Compared to general-purpose
computers, these devices are lightweight, and have the advantage of real-time
responsiveness and low power consumption.  Such devices have also been adopted
in critical areas such as intelligent factories, health care, smart homes, and
automotive industry.

Due to the importance of embedded and IoT devices and their often
infrastructure-critical nature, they have, unfortunately, become the major
targets of various attacks \cite{beniaminigalAirVolPt, loicduflotWhatIfYou2011,
greenbergJeepHackersAre,
dipintoTRITONFirstICS2018,falliereW32StuxnetDossier2010,sansinstituteConfirmationCoordinatedAttack2016,cisaHavexICSAlert2014}.
In fact, attackers can often perform code reuse attacks against these systems by
leveraging existing vulnerabilities in the code, and launch control flow
hijacking attacks when the least privilege principle is not enforced
\cite{selianinResearchingMarvellAvastar, beniaminigalAirVolPt,
artensteinBROADPWNRemotelyCompromising2017}.
These attacks can leak private and critical information, and allow attackers to
control the whole MCU, or even the devices connected to it.

To protect embedded and IoT devices, existing work has considered a number of
techniques, including firmware analysis \cite{meraDICEAutomaticEmulation2021},
fuzzing \cite{fengP2IMScalableHardwareindependent2020}, attestation
\cite{aberaDIATDataIntegrity2019}, and compartmentalization
\cite{clementsProtectingBareMetalEmbedded2017,kimSecuringRealTimeMicrocontroller2018,
clementsACESAutomaticCompartments2018}.  Compared to other techniques,
compartmentalization avoids time and resource-consuming analysis, and can
provide customized configurations for both verification and protection in
real-time.

Despite the benefits of compartmentalization, existing work has faced many
challenges in isolating the MCU resources for the rather monolithic firmware
structure.  For example, the Cortex-M architecture does not provide an address
translation mechanism such as the Memory Management Unit (MMU).  Instead, it
provides the MPU with limited functionality to divide and protect the address
space.
To offer stronger isolation, ARM has promoted TrustZone-based hardware
solutions~\cite{armlimitedPSASecurityModel2020}. Unfortunately, though,
TrustZone has yet very limited availability on existing MCU devices as described
in our survey in appendix \ref{sec:survey}.

A more concerning issue is those existing solutions---despite DMA's ubiquitous
support and heavy usage in modern MCUs---have totally ignored DMA compartmentalization
(\cite{kimSecuringRealTimeMicrocontroller2018,clementsProtectingBareMetalEmbedded2017,
clementsACESAutomaticCompartments2018}), provide partial solutions
(\cite{TockOSDesign, armARMMBEDUvisor2021}), or propose hardware modifications
(\cite{sensaouiToubkalFlexibleEfficient2019}) not available beyond the academic
boundaries. Extending current MPU-based compartmentalization solutions to
support DMA is not a trivial task. That is, DMA transfer compartmentalization is
challenging because the intrinsic dynamic characteristic of DMA aggravates the
imprecision of static analysis to define boundaries, permissions and the
security policy. Also, the currently available MPU is not meant to work in a
multi-master environment. Thus, supporting DMA on existing compartmentalization schemes 
requires a holistic re-design.

In this paper, we present \sysname, a systematic approach
to enable secure DMA operations for compartmentalization solutions targeting 
MCU-based devices that do not implement TrustZone, MMU, and
PCIe interfaces. \sysname adopts the concept of capabilities
to validate DMA operations, and enforces the least privilege
principle for RTOS holistically.

We used the FreeRTOS-MPU (\fmpu) as a basis to build a \sysname prototype due to
its wide adoption and growing popularity in the industry. We improved its
security with a more secure MPU region configuration, kernel extensions, and a
user-friendly security policy definition with explicit support for DMA
operations.

We evaluated \sysname's security and performance with both qualitative and
quantitative analysis.  The results show that compared with \fmpu, \sysname
leverages DMA while reducing the attack surface of embedded applications when
looking at six security metrics. Furthermore, \sysname incurred a low overhead
to kernel and peripheral operations. Enabling DMA operations by \sysname methods
reduced power consumption compared to solutions that do not support it. At the
same time, our solution maintained similar RAM and flash requirements as \fmpu.
A case study on a real-world PLC application further demonstrates \sysname's
capability in enabling high-performance DMA operations without compromising
security.

In summary, this work makes the following contributions:

\begin{itemize}
 
\item  We study and advocate the importance of supporting and protecting
DMA-capable peripherals for compartmentalization solutions due to
the partial or total lack of DMA support on existing solutions.

\item We present \sysname, a systematic approach to enable secure DMA operations
for embedded compartmentalization solutions that support high-performance and a
power efficient operation. 
    
\item We implemented a \sysname prototype for the official Cortex-M3/M4 port of
FreeRTOS enabling DMA operations while improving its security metrics.
      
\item We demonstrate that supporting DMA operations allows power and CPU usage
reduction, which is highly desirable for battery-powered embedded applications. 
        
\item We present a case study of \sysname on a real-world PLC application; we
further discuss \sysname methods generalization, its integration into other
compartmentalization solutions, and its limitations.
      
\end{itemize}

\section{Threat model and assumptions}\label{sec:threatmodel}

We consider an embedded device that has an MPU and one or more DMA-capable
peripherals. The device runs a monolithic firmware with the following defects: a
confused-deputy vulnerability \footnote{``A confused deputy is a deputy
(a program) that has been manipulated into wielding its authority
inappropriately'' \cite{millerCapabilityMythsDemolished} } in the managing code
of the \dctr, a malicious third-party software module with rogue usage of DMA
(i.e., result of a software supply chain attack), and a vulnerable read/write
primitive with access to the \dctr configuration. 
Furthermore, the firmware is compiled from
multiple first and third-party sources. We assume that first-party code can be
buggy, but not malicious, whereas third-party code can be malicious and buggy.
The code includes functionality for scheduling, DMA operations, peripheral
drivers, third-party libraries and user-space applications. In terms of
security, the firmware implements a compartmentalization scheme for memory and privilege
separation, but does not consider DMA operations.

If an attacker has access to any of the aforementioned defects, an attack would
be devastating. For example, a confused deputy vulnerability can be used to leak
sensitive data by abusing the \dctr authority to read sensitive information
(e.g., a password) from kernel space; a malicious task with rogue DMA usage can
change the configuration and state of other peripherals such as the GPIO 
\footnote{Particular acronyms used in this paper are summarized in appendix \S\ref{ap:acronyms}.}
controlling the smartlock of a door without triggering any fault; finally having
arbitrary read/write access to the \dctr configuration can compromise the whole
system because an attacker can read or write any memory area or peripheral
regardless of existing MPU protections.

The assumptions of our threat model are reasonable because several proposed
MPU-based compartmentalization schemes have ignored DMA security issues, or treat them as
an orthogonal security problem. However, we claim that not considering DMA in an
holistic way is insecure. This is because: first, firmware is a monolithic
binary that includes all the software routines---DMA operations are not the
exception; second, most of the modern MCUs populate and use a \dctr or other
DMA-capable peripheral; and third, DMA-capable peripherals can override 
MPU-based protections.

\section{Motivation}
\label{sec:motivation}

Embedded and IoT devices execute firmware and use an MCU as their central
processing unit. MCUs are low-power, resource-constrained computing units that
integrate a core processor, RAM, flash, and multiple peripherals in a single SoC
(System on a Chip).

Besides particular MCU hardware and firmware characteristics, embedded devices
require compartmentalization methods that cope with specific real-time and low
power requirements.  In this
context, the DMA is a predominant communication method that supports
high-performance and reduced power consumption for real-world IoT and embedded
applications.

\label{motivation}
\begin{figure}
    \centering
    \includegraphics[width=7.5cm]{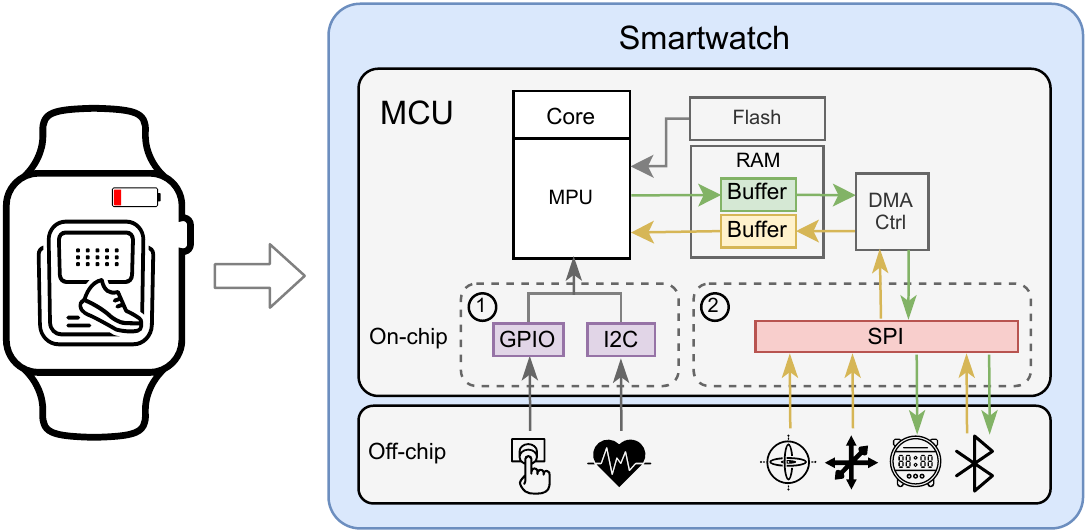}
    \caption{Architecture of an MCU-based smart watch. \Circled{1} Slow MMIO-based data flows, 
    \Circled{2} Real-time and high-throughput DMA-based data flows.}
    \label{fig:motivation}
\end{figure}

Consider the smartwatch of Figure \ref{fig:motivation}.  DMA is utilized in
high-throughput data flow such as the gyroscope, accelerometer, screen LCD, and
Bluetooth (\Circled{2} in Figure \ref{fig:motivation}).  Without DMA, these data
flows cannot assure the user experience or functionality (e.g., smooth screen
transitions or timing data acquisition for the smartwatch pedometer).  Also, the
\dctr can efficiently move data between RAM and peripherals while the core
processor is in power-saving mode.

Notice that the MPU arbiters every data flow between the core processor and the
MCU resources (\Circled{1} in Figure \ref{fig:motivation}).  This interposition
assures that the core processor accesses only the MCU resources granted by the
MPU through MMIO (Memory-Mapped I/O) operations.  On the other hand, the \dctr
directly accesses peripherals and buffers in RAM (\Circled{2} in Figure
\ref{fig:motivation}).

In this scenario, the \dctr can write and read any memory location of the MCU
without restriction. This capacity has profound security implications.  For
example, a bug in the code that manages the \dctr can wrongly command it to
write beyond the limits of the designated buffers, thus corrupting heart-beat,
distance or other critical information that the smart watch user may rely on.

\section{Background}
\label{sec:background}

In this section, we introduce essential concepts to make it easier to comprehend
the software and hardware characteristics of compartmentalization solutions for
MCUs and the challenges related to DMA.  We selected the ARM ARMv7-M
\cite{ltdARMV7MArchitecture} as our reference architecture because of its wide
adoption and popularity in IoT and embedded applications. Besides this
particularity, the concepts are generally applicable to other embedded
architectures.

\subsection{The system address map of the Cortex-M}
\label{sec:backAddressMap}

The ARMv7-M supports a single 32-bit address space. This address space is
divided into eight 0.5 GB primary partitions: code (flash), SRAM (on-chip RAM),
peripherals, two RAM regions, two device regions, and System
\cite{ltdARMV7MArchitecture}.

The ARMv7-M architecture assigns physical addresses for event entry points (vectors), 
system control and configuration.  The firmware uses these physical addresses
to access the entire memory space through Memory-mapped I/O (MMIO) or 
DMA methods. Vendors (i.e., licensees) of ARMv7-M based devices define SRAM, code
and peripheral partitions according to specific characteristics of the
MCU.

\subsection{The Memory Protection Unit}

The MPU is an optional component of the ARMv7-M architecture that implements a
scheme to protect and divide the MCU system address space into different
regions.  The MPU 
does not perform address translation to support virtual memory schemes such as
the MMU of full-fledged computers. The MPU protection
scheme enables the ARMv7-M Protected Memory System Architecture (PMSAv7) that
defines a model of privileged and unprivileged software execution
\cite{ltdARMV7MArchitecture}.

MPU regions are restricted in terms of number, size, and alignment. Usually, the
MPU is configured with eight regions, with very few exceptions on high-end MCUs
that support 16 regions \cite{stmicroelectronicsManagingMemoryProtection2020}.
The size of a region must be a power of 2, with a minimum size of 32 bytes. Each
region must be aligned naturally according to its size (e.g., a 64 bytes long
region must start at an address that is a multiple of 64). 
If regions overlap, the MPU uses the privileges of the region with the higher
number to enforce the access permissions (i.e., higher region numbers have
higher priority). Additionally, there is a background region (number -1) that,
when activated, provides access to the primary partitions of the memory map, but
from privileged software only.

\subsection{The DMA controller operation}
\label{sec:backDctrl}

The \dctr is an on-chip peripheral with master capabilities optimized to move
data from a source to a destination. As a master, the \dctr can communicate with
slaves (peripherals, RAM and flash) without intervention of the core processor,
and without interposition of the MPU. 

\begin{figure}
    \centering
    \includegraphics[width=7.5cm]{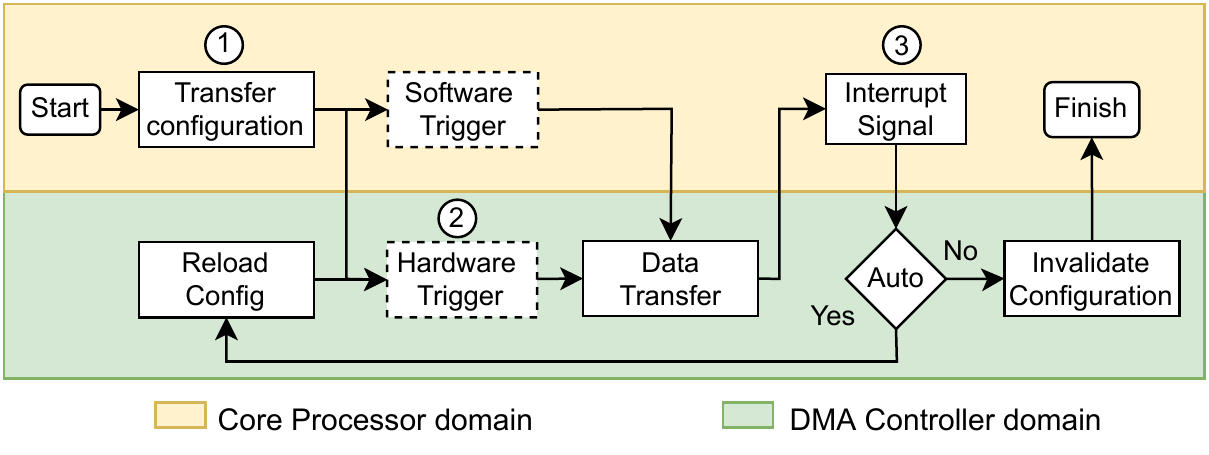}
    \caption{The 3-step DMA life-cycle including operations that are visible to the core processor and \dctr domains. Security critical operations occur in Steps 1 and 3---Adapted from \cite{meraDICEAutomaticEmulation2021}.  }
    \label{fig:lifecycle}
\end{figure}

The \dctr generally operates following a 3-step dynamic life-cycle as depicted
in Figure \ref{fig:lifecycle}.
First, the firmware configures a transfer descriptor that defines the source,
destination, and transfer size. The core processor stores the transfer
descriptors in registers mapped into the DMA controller or RAM.  Second, a
trigger issued by software or hardware signals the \dctr to start the data
transfer from the source to the destination according to the transfer
descriptor. Third, after the transfer finishes, the \dctr signals the core processor
through an interrupt, and invalidates or reloads the transfer descriptor. 

The \dctr requires addressing 
routines to work with on-chip peripherals implementing 
communication buses (e.g., SPI and I2C). The communication bus peripherals allow
the core processor to communicate with multiple off-chip peripherals (e.g., giroscope, accelerometer, 
LCD, and Bluetooth connected through SPI in Figure \ref{fig:motivation}). In a 
communication bus, the \dctr is not aware of the specific off-chip peripheral that 
is part of a particular DMA transfer. Therefore, the firmware must select the 
specific off-chip peripheral  as part of the configuration (i.e., Step 1) 
of the DMA life-cycle. A similar issue is observed in peripherals working 
as a multiplexer or proxy. For example, the ADC multiplexes various 
analog input channels and requires selecting a particular channel, or define a 
sequence of scanned channels before operating with DMA transfers.

\subsection{Open challenges}
\label{sec:backOpenChallenges}

Besides the compartmentalization efforts to improve the security of embedded
applications, there are still challenges that have not been addressed,
specifically those related to DMA, and we summarize them below:

\subsubsection{Uncertainty on protections} Current compartmentalization
solutions either overexpose regions merging physically adjacent resources, or
potentially break functionality due to static analysis imprecision. Both issues
are more concerning for DMA operations because they are intrinsically dynamic
and executed out of the core processor context. These characteristics preclude
the usage of static analysis because there is a semantic gap---due to hardware
diversity---between what is dynamically configured in a transfer descriptor and
what the \dctr performs by itself upon that configuration.

\subsubsection{Lack of holistic security solutions} Most of the MPU-based
security solutions ignore DMA operations or consider them an orthogonal issue
\cite{sensaouiIndepthStudyMPUBased2019}. However, a practical solution must
protect and leverage DMA operations to maintain a balance among security,
performance and power consumption.

\subsubsection{Hardware availability and diversity} Many compartmentalization
efforts rely on specific hardware that is not broadly available. For example,
the Platform Secure Architecture (PSA) \cite{armlimitedPSASecurityModel2020},
promoted by ARM, relies on the ARMv8-M TrustZone and a new MPU programming
model. Unfortunately, the ARMv8-M architecture is scarce in commercial MCUs.
Also, \dctrs are diverse and implement different programming models precluding
generalizations and automation.  

\subsubsection{Impractical security policy definition} Current
compartmentalization solutions require complex security policy definitions that
are not practical, and add a burden to developers.  For example, defining the
security policy for \cite{clementsACESAutomaticCompartments2018} requires
developers to define and optimize security properties in a graph traversal
algorithm. Therefore, extending this type of solutions to support DMA is not
trivial, and will make the policy definition even more complex.

\subsubsection{Backward compatibility and refactoring} Current solutions require
heavy refactoring or compiler-based procedures that can break compatibility with
legacy applications. Also, compiler-based solutions can modify the memory layout
of the generated firmware binary, which makes validating and defining the
boundaries of DMA operations challenging.

\section{System Design}
\label{sec:design}

\begin{figure}
  \centering
  \includegraphics[width=7.5cm]{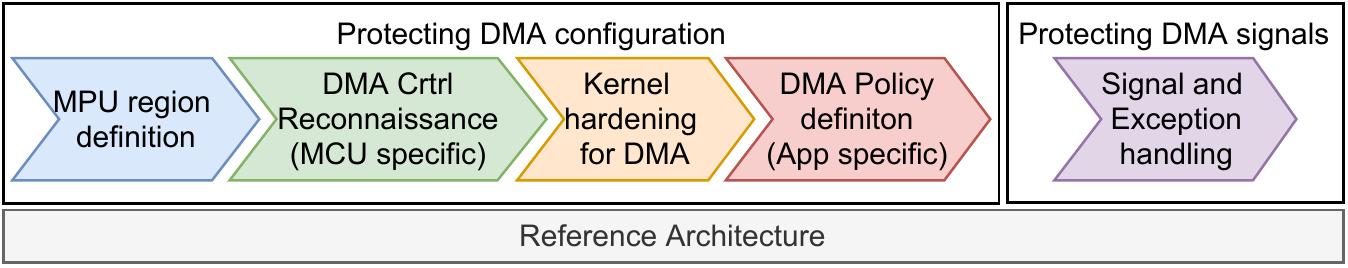}
  \caption{\sysname components including a reference architecture and a workflow
   to enable DMA operations protecting the steps one (Configuration) and three (Signals) 
   of the DMA life-cycle.}
  \label{fig:overview}
\end{figure}

\sysname is a systematic approach to enable DMA operations on
compartmentalization solutions of embedded applications.  
Our approach includes a reference architecture built on top of \fmpu,
and a workflow (Figure \ref{fig:overview}) to enable secure DMA operations. 
It addresses the challenges mentioned earlier 
with the following goals:

\begin{itemize}
  
    \item \textbf{Explicit protections:} \sysname should define
    explicit resources to maintain application functionality and
    avoid uncertainty.
      
    \item  \textbf{Holistic DMA support:} \sysname should support 
    DMA operations as an intrinsic characteristic of the compartmentalization schema.
      
    \item \textbf{Power and performance:} \sysname should respect the
      power, performance and timing constraints required by embedded
      and real-time applications.

    \item  \textbf{Compatibility:} \sysname should rely exclusively on
    commonly-available hardware of MCUs, and it should consider its 
    limitations and diversity.

    \item  \textbf{Applicability:} \sysname should be pragmatic in terms of
    security policy definition and usage.
           
    \item \textbf{Backward support:} \sysname should be amenable with legacy
    applications requiring little or no engineering effort to support them.

\end{itemize}

\subsection{\sysname reference architecture}

\begin{figure}
  \centering
  \includegraphics[width=8.5cm]{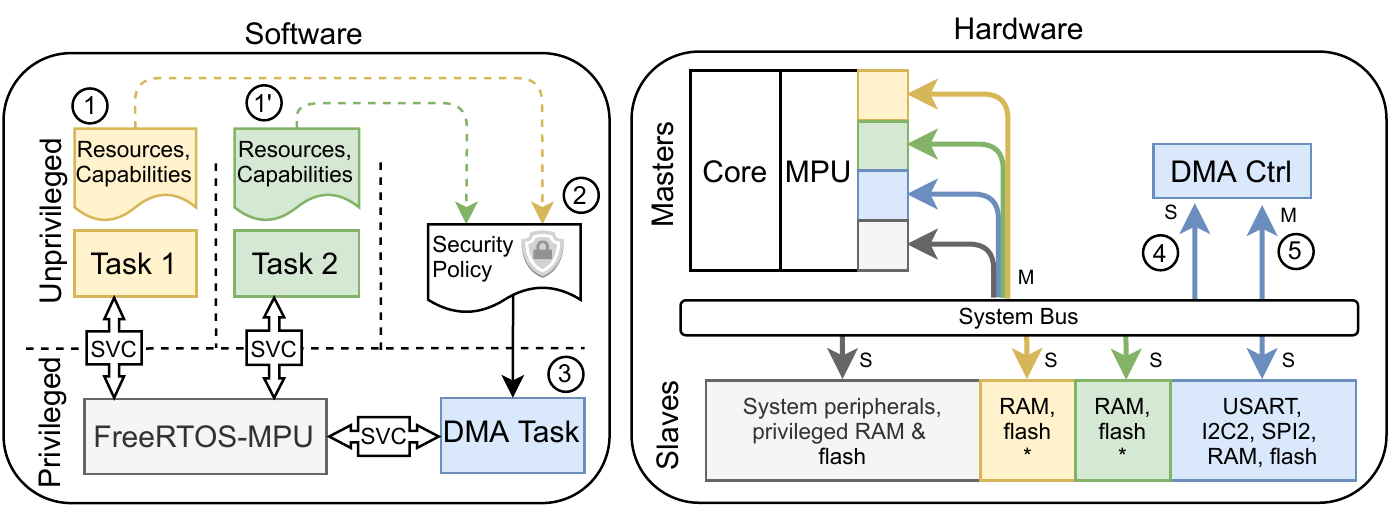}
  \caption{\sysname reference architecture including a software and hardware perspective. 
  ``M" and ``S" denote master and slave interfaces of the Cortex-M respectively. 
  (*) denotes user-defined regions mapped to MCU resources.}
  \label{fig:design}
\end{figure}

\sysname defines a task as the unit of compartmentalization.  A
\emph{task} is a natural partitioning scheme that the developer explicitly
defines with the required resources and capabilities (\Circled{1} and
\Circled{1'} in Figure \ref{fig:design} ) to implement specific functionality.
\sysname uses the security policy (\Circled{2} in Figure \ref{fig:design} ) to
securely override the MPU protections through a trusted DMA task (\Circled{3} in
Figure \ref{fig:design}) that configures the \dctr upon a request. \sysname
enforces a role separation, assuring that no single task has enough privileges
to configure the \dctr and control the data of a DMA transfer simultaneously.
This means that the DMA task can control from \emph{where} the \dctr reads or
writes, while the user tasks (Task 1 and Task 2 in Figure \ref{fig:design})
control \emph{what} to read or write within their compartmentalized resources.

Our proposed architecture separates the kernel and the DMA task (i.e., they run
on different threads) because of two
design considerations: first, it maintains the microkernel architecture of \fmpu
with its intrinsic compartmentalization; and second, it 
makes the solution more generic and
extensible by developers. The second reason is essential because the DMA task
must implement the drivers for a particular \dctr, which is not generic. Also,
implementing the DMA task functionality directly in the kernel will
require low-level development knowledge and precludes the mitigation of the security
risk associated with a larger trusted computing base (TCB). We will extend 
our discussion of the TCB in \S\ref{sec:discussion}.

The downside of this separation is a small overhead due to the extra SVC
(SuperVisor Call) to communicate the DMA task and the kernel. Regardless of 
privilege level, the SVC is mandatory to keep a thread-safe intercommunication 
through FreeRTOS primitives. Nevertheless, we
demonstrate in \S\ref{sec:evaluation} that the overall performance of
our solution is adequate for embedded applications.

\subsection{\sysname MPU region definition}
\label{sec:mpuregions}

D-Box uses the \fmpu explicit definition of MPU regions to avoid the inaccuracy
and incompleteness of static and dynamic analysis used by other solutions. 
The explicit definitions provide certainty about the resources and permissions 
that tasks access during execution time. This characteristic maintains task’s functionality, assures 
that the \dctr configuration is not inadvertently exposed,  
and provides the means to define a deterministic security policy.

The standard \fmpu MPU region definitions are too permissive. 
\sysname redefines these regions, as depicted in Figure \ref{fig:regionsys},
to implement a stricter compartmentalization scheme that is compatible with DMA operations.
The new MPU region definitions and its security properties are as follow:

\subsubsection{Background region (-1)}  This region assures that unprivileged
code has no access to any MCU region by default. Specifically, this region
protects any DMA-capable peripheral that might be exposed by the former region
number 3. 

\subsubsection{Syscalls region (0)} This region grants unprivileged code access
to valid syscalls entry points (i.e., SVC in Figure \ref{fig:design}).  The
kernel uses it to avoid task code from asking elevation of privileges from
arbitrary code locations---a common symptom of a control flow hijacking attack.
This region is similar to the former region number 0 but does not include the
task code.

\subsubsection{Task code region (1)} This region isolates the task code and
reduces gadgets for code-reuse attacks.  It avoids an attacker controlling a
task to reuse artifacts that could diverge DMA operations identified on other
task codes. Using this region might require considerable refactoring. Developers
will need to allocate all code and constants accessed during execution under a
single MPU region. However, this is optional, and developers can define region 1
as the former region 0 to maintain backward compatibility and reduce engineering
effort, as per our design goals.

\subsubsection{Task stack region (2)} This region is not executable to avoid
code injection attacks. It also isolates the stack of a task from other memory
areas. It detects and prevents out-of-bounds read or write operations. This
protection restricts attacks or crashes within the boundaries of a single task.
The functionality and properties are similar to the former region 4.  Also, this
region delineates a valid source or destination for DMA operations.

\subsubsection{User-defined regions (3), (4), (5)} These regions allow
developers to grant or deny task access to
peripheral, RAM, and flash similar to the former regions 5, 6, and 7. 
D-Box's new region numbering schema prevents user-defined regions from
overriding kernel code and memory regions because of higher region number
precedence. Simultaneously, the user-defined 
regions can override the task code, stack, and syscalls regions to deny access 
to specific sub-regions. For example, a task can initialize variables in the stack 
or access syscalls only during initialization, and later, use the user-defined 
regions to grant read-only permission to initialized variables 
and block access to syscalls---this is a characteristic already supported by \fmpu.
Similar to the stack region, user-defined regions delineate valid
source and destination locations for DMA operations.

\subsubsection{Kernel Code, and stack and heap regions (6, and 7)} These regions
protect kernel code and memory from unprivileged access. The kernel's higher
region numbers avoid lower-numbered regions to override critical sections---an
evident defect in the former schema. The higher region numbers allow keeping
immutable task identification data in kernel space to support a capability
validation schema of DMA operations that we discuss in \S
\ref{sec:securitypolicy}.

\begin{figure}
    \centering
    \includegraphics[width=8.5cm]{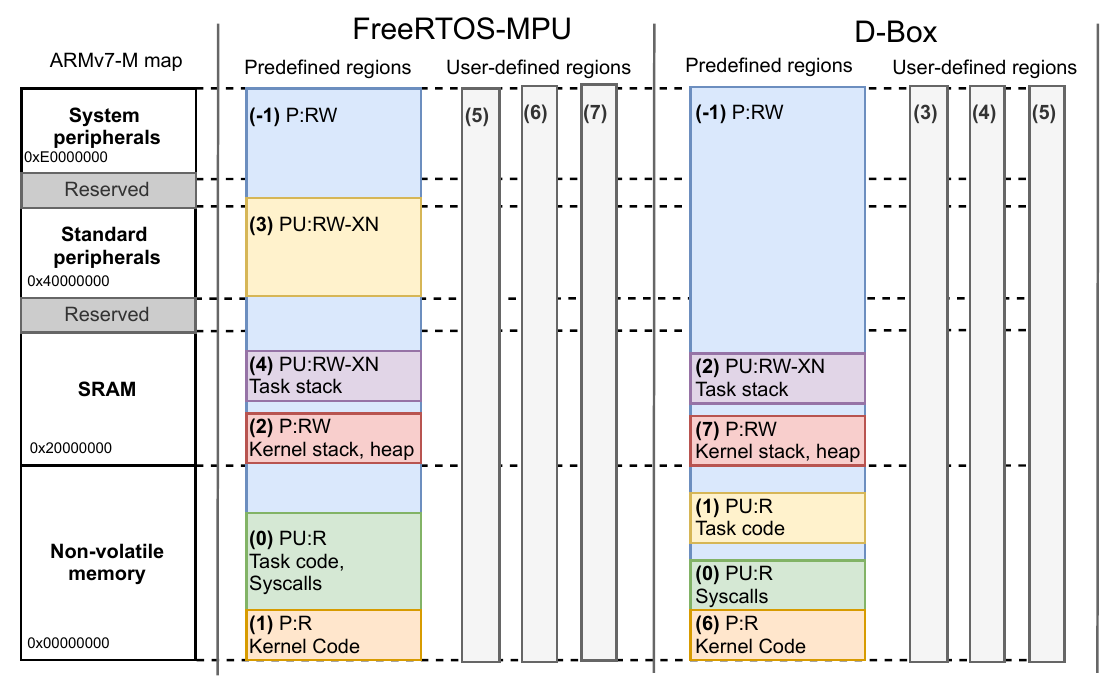}
    \caption{MPU region definitions for \fmpu and \sysname. Execution levels: (\textbf{P})rivileged, and (\textbf{U})nprivileged. 
    Permissions: (\textbf{R})ead, (\textbf{W})rite, and Execute Never(\textbf{XN}). The highest region number takes priority
    when regions overlap.}
    \label{fig:regionsys}
\end{figure}

\subsection{\dctr reconnaissance}
\label{sec:reconnaissance}

\sysname includes as part of its design a manual reconnaissance phase of the
\dctr to determine its programming model, number of DMA channels, and critical
areas where the main core stores the \dctr configuration.  The location is
specific to each MCU and might include RAM and peripheral MMIO areas.

The reason to include this phase is the diversity of MCU hardware---an issue
largely described by the re-hosting research community
\cite{fengP2IMScalableHardwareindependent2020,meraDICEAutomaticEmulation2021}.
This diversity precludes the use of compiler-based compartmentalization
solutions, or secure programming languages that simply fall short without a
manual analysis of the MCU hardware.  Moreover, the \dctr is not the only DMA
capable peripheral. Modern MCUs populate other peripherals that support DMA by
themselves, for example, 2D graphic accelerators
\cite{stmicroelectronicsChromARTAccelerator}, USB, and CAN bus controllers.  All
of these peripherals require reconnaissance to keep the properties of the
compartmentalization solution.

\subsection{Kernel hardening to support DMA operations}
\label{sec:hardening}

\sysname defines DMA protection rules (extensions) that the kernel must enforce
during the creation of tasks, and the allocation of memory for the transfer 
descriptors of the \dctr. These rules cannot be implemented 
in the MPU because the kernel re-configures the MPU on each context switch 
replacing the configuration used by the last task with the 
configuration of the currently scheduled task. The general protection 
rules are as follows:

\begin{itemize}

    \item User-defined regions cannot be mapped to the \dctr slave
       interface because it is used for configuration.
       
     \item Transfer descriptors must be protected either by the previous rule 
       or by keeping these data structures in kernel space.
       
     \item The configuration of the \dctr is only performed by the
       trusted DMA task upon a request and a policy verification.
       
     \item The DMA task cannot access the stack of other tasks or
       user-defined regions.
       
     \item The user-defined regions cannot be mapped to other task
       stacks.
       
 \end{itemize}

The last two rules complement the
\sysname MPU region definitions to avoid user-defined regions from exposing the
stack of other tasks.

\subsection{DMA policy definition}
\label{sec:securitypolicy}

\sysname defines a capability-based security model \cite{hardyConfusedDeputyWhy1988} for DMA operations. 
A capability is an immutable  reference to an object (RAM, peripherals, or flash)
associated with access rights (i.e., read, and write). The possession of a capability grants the 
owner (i.e., a task) the defined right to interact with the object.

We selected the capability-based model because it can mitigate confused deputy
vulnerabilities by design \cite{millerCapabilityMythsDemolished}---a known
limitation of ACL-based security models.  This type of vulnerabilities are
difficult to mitigate for current compartmentalization solutions
\cite{clementsACESAutomaticCompartments2018} because of the uncertainty of
resources exposed to each compartment.  However, \sysname uses explicit
definition of resources which supports a capability-based policy for DMA
operations on peripherals.

\sysname defines DMA capabilities as a combination of an on-chip peripheral
(i.e., the object) and \textbf{E}xtensible \textbf{A}ccess \textbf{R}ights
(EAR). The EAR includes the
standard read and write permissions, and optional parameters to support the
addressing schema to select off-chip peripherals, as we described in \S
\ref{sec:backDctrl}.

\sysname enforces the DMA policy by verifying the source and destination 
for read or write operations. For a write operation, \sysname verifies 
that the source (always a buffer) is contained in the task stack or in 
any user-defined region, and that the destination (always a peripheral) 
is defined in the capability with the corresponding permission (i.e., write).  
Similarly, for a read operation, \sysname verifies that the destination (always a
buffer) is contained in the task stack or any user-defined region, and
that the source (always a peripheral) is defined in the capability with
the corresponding permission (i.e., read).  For {\tt FullDuplex}
operations, \sysname combines the verification of read and write
operations simultaneously. 

Notice that a capability does not explicitly define memory ranges
because \sysname uses the---already explicit---task MPU 
configurations of the stack and user-defined regions for this purpose. This 
characteristic assures that the task has read or write access to peripherals 
through DMA, and that the memory buffers associated with the DMA operations 
are also accessible to the task with proper permissions granted by the MPU. 
Additionally, the DMA capabilities do not require MPU region re-definition 
to access the associated peripheral because the \dctr can override the MPU protections. 
This property allows the definition of an extensive security 
policy with more flexibility that is aware of, but not limited 
by the number of available MPU regions.

\subsection{Signal and exception handling} \sysname handles the signals when a
DMA transfer finishes, and the events when a security violations is detected.

\subsubsection{Signal registration and notification} \sysname defines an
automatic method to register a task for signals or notifications derived from
the \dctr interrupts (i.e., step 3 of the DMA life-cycle).  The registration
occurs after the verification of the security policy for a DMA request. Since
the capability-based policy provides certainty about the requester (i.e., a
task), \sysname registers the task exclusively to receive a notification when a
particular DMA transfer finishes. The notification mechanism manages the 
Interrupt Service Routine (ISR) of the \dctr with a minimum code base running in
privileged mode. After managing the interrupt, \sysname mechanism notifies the
task using the SVC API and drops its privileges.  The number of
concurrently-registered tasks to receive notifications depends on the number of
DMA channels supported by the \dctr. Technically speaking, each DMA channel
serves a single request associated with a unique task.  \sysname assures proper
resource management according to the characteristics of the \dctr that were
obtained by the reconnaissance phase. 

Since FreeRTOS-MPU does not provide any specific interrupt management
system, developers can implement notification mechanisms similar to
\sysname for other system interrupts.  The primary consideration is that the
core processor always executes the ISR in privileged level. Therefore,
the ISR has to be trusted and its execution deprived of privileges
before returning control to the tasks.

\subsubsection{ Exception handling} \sysname exception handling contains faulty
operations in the boundaries of a task without affecting other task operations.
\sysname considers three types of exceptions: faulty DMA requests, overlapped
user-defined regions on stack or \dctr, and MPU region violations. Only the MPU
region violation will trigger a hardware exception---the other two are managed
entirely in software.

The first type of exception occurs during the validation of the DMA request
parameters. If the request violates the security policy, \sysname rejects the
request and notifies the requester. The requester task should implement a method
to handle the rejection notification, and continue its execution.
The second type of exception can occur during initialization/creation
of a task, or during its execution when a task requests to redefine
the MPU regions. If the task is already running, \sysname ignores the
request to redefine the MPU regions, returns an error message, and
lets the task continue its execution. If the task is not running,
\sysname voids the initialization/creation of the task and clears 
it from the scheduler.
The third type of exception occurs because of mismatches or permission
violations of the MPU regions.  \sysname considers this type of exception
severe. Hence, it will stop the offending task, and remove it from the
scheduler, keeping the rest of task running.

\section{Implementation}
\label{sec:implementation}

We implemented a \sysname prototype for the official Cortex-M3/M4 port of
FreeRTOS version 10.4.1. Our implementation is divided into kernel 
hardening and extensions, and a privileged task that manages the DMA 
transfers to override the MPU protections
according to the security policy of \sysname.

In this section, we refer to the core FreeRTOS kernel and the port layer for the
Cortex-M as the ``kernel'' code. Also, we refer to the MCU vendor-specific code
as the ``drivers''. Our prototype targets the popular STM32 MCU family of ST
Microelectronics for the vendor-specific sections.  However, the kernel changes,
API, and functions are vendor-agnostic, and compatible with any Cortex-M3/M4
MCU. 

Our complete prototype includes three lines of assembly language (for the
highly-optimized context switch routine), 1200 lines of C code for the DMA task,
drivers and ISR, and 220 lines of C code for the kernel modifications.

\subsection{Kernel hardening and extensions}

We modified the \fmpu kernel to support an extra predefined MPU region per task
according to region number 1 in Figure \ref{fig:regionsys}.  With this
modification, each execution of the context switch routine configures the MPU to
grant access to the task stack, task code, and the three user-defined regions.
The regions for the kernel data, kernel code and syscalls are configured once,
and maintained during the entire firmware execution. 

We extended the structure of the Cortex-M Thread Control Block (TCB) to store a
pointer to an array of capabilities for the security policy.  This array is a
structure passed as a parameter during the initialization of the tasks.
Every capability entry in the array contains a peripheral ID (i.e., the physical
address of peripheral), a bit field for granted rights flags (e.g., Read, Write,
FullDuplex), and an option field for the off-chip addressing schema, which is
peripheral-specific.

\lstset{
    numbers=left,
    xleftmargin=3.5em,
    frame=single,
    framexleftmargin=2.5em,
    stepnumber=1,
    language=C,
    showstringspaces=false,
    basicstyle=\tiny,
    keywordstyle=\bfseries\color{green!40!black},
    commentstyle=\itshape\color{purple!40!black},
    identifierstyle=\color{blue},
    stringstyle=\color{orange},
    morekeywords={eRead, eWrite, eFullDuplex},
    caption=Capabilities defined in C code structures,
    label=listing:Permission,
}

\begin{lstlisting}[frame=single] 
static const PeripheralPermission_t xPermission[portTOTAL_NUM_PERMISSIONS] = 
{ 
  {(uint32_t *)SPI1,  (eRead | eWrite | eFullDuplex), SS_FRAM_ }, 
  {(uint32_t *)I2C2,  (eRead | eWrite | eFullDuplex), 0x08     }, 
  {(uint32_t *)ADC1,  (eRead),(ADC_CHANNEL_0 | ADC_CHANNEL_4)  } 
};
\end{lstlisting}

We used simple C language structures to define the capabilities
because \fmpu uses similar structures for the user-defined MPU regions. Hence,
developers are familiar with them.  Also, this scheme saves memory resources
without requiring a parsing routine for more sophisticated formats (e.g., json)
that would increase the footprint of the solution.
    
We defined all the kernel changes and extensions with pre-processor C directives
to allow the activation or deactivation of \sysname in the original FreeRTOS
configuration file.  Our changes are also backward compatible with applications
developed for the standard \fmpu.  For example, developers can choose to use
exclusively the user-defined regions to access peripherals and the required
memory areas according to the restrictions implemented in the kernel.

\begin{figure}
  \centering
  \includegraphics[width=8cm]{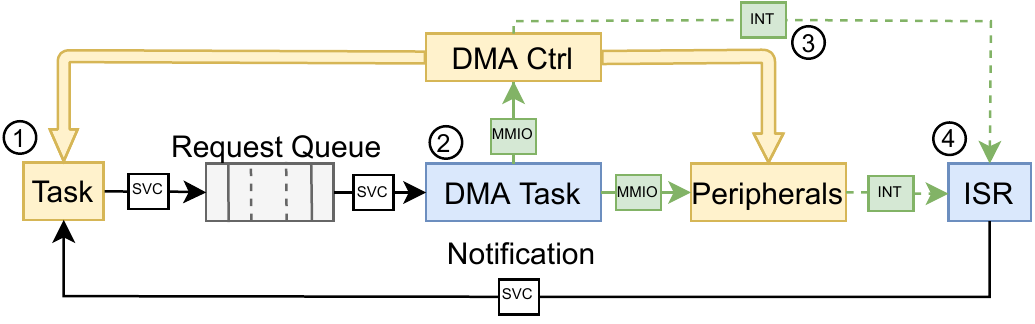}
  \caption{ \sysname implementation including the data flows, the independent channel created by the \dctr, and the notification mechanism.}
  \label{fig:dataflow}
\end{figure}

\subsection{The DMA task and data flow}

The DMA task (\Circled{2} in Figure \ref{fig:dataflow}) is a trusted privileged
task that manages DMA requests from other tasks (\Circled{1} in Figure
\ref{fig:dataflow}). The main functions of the DMA task include the validation
of the DMA request, the configuration of peripherals including the \dctr, and
the registration of the requesting task for the notifications.  The ISR handles
the interrupts from the \dctr and other peripherals (\Circled{3} and \Circled{4}
in Figure 6) and delivers the notification to the requesting task.

The DMA task and the ISR use exclusively standard FreeRTOS primitives (i.e.,
queues, and notifications for Inter-Process Communication (IPC) through the
FreeRTOS SVC interface). This implementation decision assures two properties:
First, the IPC is thread-safe. Second, developers can extend \sysname with more
peripheral drivers and functionalities without the knowledge of low-level kernel
development.

\section{Evaluation}
\label{sec:evaluation}

\addtolength{\tabcolsep}{-3.5pt}

We evaluated \sysname to answer the following questions:

\begin{enumerate}
  
\item What is the attack surface of an application that supports DMA through
\sysname procedures? How does our solution compare to \fmpu and other similar
solutions?
      
\item What is the impact of \sysname in terms of performance, memory usage, and
power consumption for embedded applications?
      
\item What are the security and operational benefits of \sysname for real-world
embedded applications?
      
\end{enumerate}

To answer Question 1), we performed a qualitative and quantitative security
analysis of \sysname in \S\ref{sec:SecurityAnalysis}. For Question 2), we
conducted performance, memory, and power analyses in
\S\ref{sec:PerformancelAnalysis}, \S\ref{sec:memoryanalysis}, and
\S\ref{sec:powerenergy} , respectively.  Finally, we answer Question 3) and
present in \S\ref{sec:caseStudies} an end-to-end case study that reveals a
DMA-enabled compartmentalization solution with negligible overhead and reduced
power requirements for real-world applications.

Besides the specificities of each subsection, the common characteristics of our
setup environment are as follows:

\begin{itemize}
  
\item Development board: ST NUCLEO-L152RE.
  
\item MCU: STM32L152RE Cortex-M3@32Mhz, 512kB flash, 80kB RAM, 154kB peripheral
MMIO, 2 \dctrs.
  
\item OS: FreeRTOS 10.4.1

\item MPU: 8 regions, 32 bytes minimum region size.
  
\item Peripheral operations: read and write operations on I2C, USART, SPI and ADC peripherals.
  
\end{itemize}

\subsection{Security analysis}
\label{sec:SecurityAnalysis}

\subsubsection{Security metrics}
\label{sec:metrics}

Based on the least-privilege design goal, well-known defense mechanisms, and the
security framework BenchIoT \cite{almakhdhubBenchIoTSecurityBenchmark2019}, we
defined the following quantitative and qualitative metrics to assess the
security properties of \sysname:

\begin{itemize}
  
\item \textbf{Memory region ratio:} is a metric that measures the effectiveness
of the compartmentalization of MCU resources by computing the ratio of the size of memory
areas exposed during a task's execution slot to the total size of each memory
area of the MCU (i.e., RAM), flash and peripherals. A lower ratio represents a
better compartmentalization.
    
\item \textbf{Number of ROP gadgets:} this metric computes the number of code
snippets exposed in flash for ROP (i.e., return oriented-programming) attacks
during a task's execution slot.  A lower number of ROP gadgets reduces the
attacker's capability to hijack task execution, and perform arbitrary actions on
the MCU.
    
\item \textbf{Data execution prevention:} is a security mechanism that enforces
the W$\oplus$X principle.  This mechanism precludes the execution of data
(payloads) controlled by an attacker.
      
\item \textbf{Code execution level segregation:} is a security mechanism that
differentiates and limits the access to system-critical resources.
      
\item \textbf{Stack protection:} is a security mechanism that detects and
prevents operations outside the boundaries of the task stack. This mechanism
prevents a faulty or compromised task from writing or reading beyond the limits
of the task's stack, but it does not detect stack corruption within the valid
boundaries.
    
\item \textbf{Extensible Access Rights (EAR):} is a security mechanism that
allows defining security policies for off-chip peripherals for DMA operations.
      
\end{itemize}

\subsubsection{Quantitative security analysis}

In this section, we analyzed quantitatively the memory region ratio and the
number of ROP gadgets for an unprivileged task (i.e., user task) running on
\fmpu and \sysname. We considered the regions and subregions of flash, RAM and
peripherals, and measured the standard and the worst-case scenarios for each
tool.  The standard scenario corresponds to a configuration that does not use
any of the MPU user-defined regions, whereas the worst-case uses the MPU
user-defined regions to expose all possible memory regions to the user task
(i.e., maximum exposure).

For the analysis of the number of ROP gadgets, we further divided the flash
region into syscalls, kernel, drivers, libc, DMA routines and the user task. To
identify the ROP gadgets, we used Ropper \cite{sashRopperRopGadget2014}.  To map
the location of ROP gadgets with the code subregions, we used the reverse
engineering tool Ghidra \cite{nsaGhidra}.

The result of our analysis (Table \ref{table:memoryexposure}) demonstrates that
the standard configuration of \sysname can reduce the region ratio to
0\% and 1\% for standard peripherals and user space flash respectively,
compared to \fmpu. Also, \sysname assures (either by MPU region number
precedence, API filtering, or security policy) that the kernel, the \dctr, and
the DMA task's stack are always protected as security critical regions.
Conversely, the analysis result exposes a faulty design of the original \fmpu
that allows exposing 100\% of the kernel to unprivileged tasks through
user-defined regions.  It is worth noting that accessing system peripherals
always requires privileged level of execution, even when the MPU is disabled or
unavailable \cite{ltdARMV7MArchitecture}.  This is reflected in our results
showing both solutions blocking the access to system peripherals regardless of
the MPU user-defined regions.

\begin{table}[]
    \scriptsize 
    \centering
    \begin{tabular}{cl|ll|ll}
                                                               & \multicolumn{1}{c|}{} & \multicolumn{2}{c|}{\textbf{\fmpu}}     & \multicolumn{2}{c}{\textbf{\sysname}} \\ \hline
    \multicolumn{1}{l|}{\textbf{Region}}                       & \textbf{Subregion}    & \textbf{Std. {[}\%{]}} & \textbf{W-C {[}\%{]}} & \textbf{Std. {[}\%{]}} & \textbf{W-C {[}\%{]}} \\ \hline
    \multicolumn{1}{l|}{\multirow{3}{*}{\textbf{Flash}}}       & Kernel                & 0                     & 100                   & 0                     & 0                     \\
    \multicolumn{1}{l|}{}                                      & Syscalls              & 100                   & 100                   & 100                   & 100                   \\ 
    \multicolumn{1}{l|}{}                                      & User space            & 100                   & 100                   & 1                     & 100                   \\ \hline
    \multicolumn{1}{l|}{\multirow{2}{*}{\textbf{RAM}}}         & Kernel                & 0                     & 100                   & 0                     & 0                     \\
    \multicolumn{1}{l|}{}                                      & User space            & 6.5                   & 100                   & 6.5                   & 87                    \\ \hline
    \multicolumn{1}{l|}{\multirow{3}{*}{\textbf{Peripherals}}} & Sys. periph.          & 0                     & 0                     & 0                     & 0                     \\
    \multicolumn{1}{l|}{}                                      & Std. periph.          & 100                   & 100                   & 0                     & 98.7                  \\
    \multicolumn{1}{l|}{}                                      & \dctr                 & yes                   & yes                   & no                    & no                   
    \end{tabular}
    \caption{Memory region ratio for Standard (\textbf{Std.}) and worst-case (\textbf{W-C})  configurations of an unprivileged task running on \fmpu and \sysname. Lower values represent a better protection according to the least-privilege principle. }
    \label{table:memoryexposure}
\end{table}

\sysname exposes only 2.4\% (13) of the total number (541) of ROP gadgets for
its standard configuration -- i.e., 41 times less than \fmpu (Table
\ref{table:ropexposure}). The reason for this drastic difference is the absence
of drivers in user space because \sysname accesses peripherals through the
kernel primitives (syscalls) and the DMA task that has access to the drivers.
Also, \sysname provides the MPU region number 1 to grant access only to the
task's code -- which does not include libc -- the highest contributor of ROP
gadgets according to our analysis. However, we consider that libc is an standard
resource used by tasks on real embedded applications. Adding libc to the
standard \sysname configuration will expose 71,7\% of ROP gadgets, which is
still 24.4\% less than \fmpu.

\begin{table}[]
    \scriptsize 
    \centering
    \begin{tabular}{l|lll|lll}
                       & \multicolumn{3}{c|}{\textbf{\fmpu}}                         & \multicolumn{3}{c}{\textbf{\sysname}}                       \\ \hline
    \textbf{Location}  & \textbf{Avl. [\#]} & \textbf{Std. [\#]} & \textbf{W-C [\#]} & \textbf{Avl. [\#]} & \textbf{Std. [\#]} & \textbf{W-C [\#]} \\ \hline
    \textbf{Syscalls}  & 7                  & 7                  & 7                 & 7                  & 7                  & 7                 \\
    \textbf{Kernel}    & 11                 & 0                  & 11                & 11                 & 0                  & 0                 \\
    \textbf{Drivers}   & 142                & 142                & 142               & 142                & 0                  & 142               \\
    \textbf{Libc}      & 365                & 365                & 365               & 365                & 0                  & 365               \\
    \textbf{DMA task}  & n/a                & n/a                & n/a                & 10                 & 0                  & 10                \\
    \textbf{User task} & 6                  & 6                  & 6                 & 6                  & 6                  & 6                 \\ \hline
    \textbf{Total}     & 531                & 520 (98\%)         & 531 (100\%)       & 541                & 13 (2.4\%)         & 530 (98\%)       
    \end{tabular}
    \caption{Number of ROP gadgets available (\textbf{Avl.}) and exposed on specific sections of the flash
    for standard (\textbf{Std.}) and worst-case (\textbf{W-C}) scenarios. A lower number of gadgets 
    represents a better protection and fewer chances of code-reuse attacks.}
    \label{table:ropexposure}
\end{table}

\subsubsection{Qualitative Security Analysis}

\sysname inherits many security properties from \fmpu, and it adds distinctive
protections to support DMA.  As described in Table \ref{table:seccomparisson},
the DMA protection is also partially supported by TockOS and uVisor, whereas the
capabilities with EAR for off-chip peripherals is only supported by \sysname.

\begin{table}[]
    \scriptsize 
    \centering
    \begin{tabular}{lccccccc}
    \multicolumn{1}{l|}{\textbf{Security feature}}                           & \multicolumn{1}{l}{\textbf{uVisor}} & \multicolumn{1}{l}{\textbf{TockOS}} & \multicolumn{1}{l}{\textbf{EPOXY}} & \multicolumn{1}{l}{\textbf{MINION}} & \multicolumn{1}{l}{\textbf{ACES}} & \multicolumn{1}{l}{\textbf{\fmpu}} & \multicolumn{1}{l}{\textbf{\sysname}} \\ \hline
    \multicolumn{1}{l|}{\textbf{Code isolation}}    & o                                   & x                                   & o                                  & x                                   & x                                 & o                                  & xo                                    \\
    \multicolumn{1}{l|}{\textbf{Data isolation}}    & x                                   & x                                   & o                                  & x                                   & x                                 & x                                  & x                                     \\
    \multicolumn{1}{l|}{\textbf{Periph. isolation}} & x                                   & xo                                  & o                                  & x                                   & x                                 & x                                  & x                                     \\
    \multicolumn{1}{l|}{\textbf{DMA protection}}    & xo                                  & xo                                  & o                                  & o                                   & o                                 & o                                  & x                                     \\
    \multicolumn{1}{l|}{\textbf{DEP}}               & xo                                  & x                                   & x                                  & x                                   & x                                 & x                                  & x                                     \\
    \multicolumn{1}{l|}{\textbf{Exec. level seg.}}  & x                                   & x                                   & x                                  & x                                   & x                                 & x                                  & x                                     \\
    \multicolumn{1}{l|}{\textbf{Stack protection}}  & x                                   & x                                   & x                                  & x                                   & x                                 & x                                  & x                                     \\
    \multicolumn{1}{l|}{\textbf{Capability EAR}}      & o                                   & o                                   & o                                  & o                                   & o                                 & o                                  & x                                     \\ \hline
    \multicolumn{8}{c}{x = Yes, o = NO, xo = partial/optional}                                                                                                                                                                                                                                                             
    \end{tabular}
    \caption{Comparison of security features supported by \sysname and related tools.}
    \label{table:seccomparisson}
\end{table}

\sysname keeps each task stack in an independent MPU region during its
execution. This configuration detects and prevents buffer overflows that may
affect surrounding memory areas, but it cannot prevent stack corruptions within
the MPU region.  The same consideration is maintained for the DMA operations. In
this case, a wrong or malicious DMA transfer request can corrupt the stack of
the requesting task or the user-defined regions, but \sysname's immutable
capability-based policy assures that the DMA transfer will only affect the
requesting task resources.  This characteristic, accompanied by the explicit
declaration of resources, mitigates the firmware defects described in our threat
model, even when the attacker controls the code of a task.

Finally, \sysname supports the protection of DMA with EAR for I2C, SPI and ADC
peripherals.  This characteristic extends the granularity of the security policy
beyond the boundaries of on-chip peripherals, which is not supported by other
solutions at all.  \sysname does not overload the DMA operation. Rather, it
simply enforces the security policy on mandatory configuration parameters that,
otherwise, are written on the \dctr, or managed by critical routines without a
systematic verification.

\subsection{Performance analysis}
\label{sec:PerformancelAnalysis}

In this section, we present micro and macro benchmarking used to measure
\sysname's overhead on kernel and common peripheral operations, respectively.

\subsubsection{Micro-benchmarking}
\label{sec:MicroBenchmark}

This analysis shows the overhead of \sysname on specific kernel operations,
including the context switch routine, task creation syscall, and DMA validation.

\textbf{Setup:} we measured the overhead introduced by \sysname using the
Cortex-M's cycle count register of the Data Watch Point and Trace Unit (DWT)
\cite{armCortexM4TechnicalReference}. 
We manually added instrumentation to obtain the number of cycles used in each
operation. We took 10 samples and averaged the number of cycles for each
operation.

\textbf{Results:} \sysname adds a reasonably low overhead to the context switch
routine (i.e., 1.74\% as described in Table \ref{table:microbench}) due to the
extra MPU region (Task code in Figure \ref{fig:regionsys}) modified on each
context switch.  Also, the creation of tasks shows a linear overhead O(n)
because of the validation of each task’s stack and user-defined regions against
previously-created tasks, and the \dctr regions. This validation occurs once
during task creation, and does not affect the task at execution time. 
\sysname uses on average 657 cycles for the DMA transfer validation. This
validation overhead differs between peripherals and operations
because of the EAR support for off-chip peripherals. We observed  
the higher overhead on the I2C operations that need the 
addressing schema of EAR, and the lower overhead on USART operations that do not 
use this schema.

\begin{table}[]
    \scriptsize 
    \centering
    \begin{tabular}{l|lll}
    \multirow{2}{*}               & \textbf{\fmpu}        & \textbf{\sysname }       & \textbf{Overhead}  \\ \hline  
    \textbf{Parameter}            & \textbf{[\# cycles]}  & \textbf{[\# cycles]}     & \textbf{[\%]}      \\ \hline     
    \textbf{Context switch}       & 287                   & 292                      & 1.74               \\
    \textbf{Task creation 0}      & 12730                 & 13047                    & 2.49               \\
    \textbf{Task creation 1}      & 12730                 & 13332                    & 4.73               \\
    \textbf{Task creation 2}      & 12730                 & 13617                    & 6.97               \\
    \textbf{Task creation 3}      & 12730                 & 13902                    & 9.21               \\
    \textbf{DMA validation Min.}  & n/a                   & 569                      & n/a                \\
    \textbf{DMA validation Avg.}  & n/a                   & 657                      & n/a                \\
    \textbf{DMA validation Max.}  & n/a                   & 792                      & n/a               
    \end{tabular}
    \caption{\sysname micro-benchmarking for context switch, task creation and DMA transfer validation.}
    \label{table:microbench}
\end{table}

\subsubsection{Macro-benchmark of peripheral operations}

\label{sec:PeripheralPerformance}

\begin{figure*}
    \centering
    \includegraphics[width=18cm]{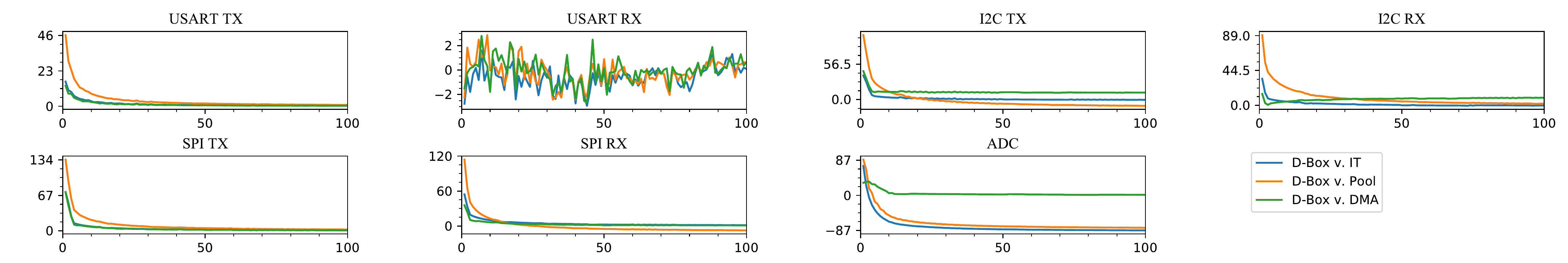}
    \caption{Performance overhead [\%] of \sysname versus Interrupt (IT), Pooling (Pool) and 
    Insecure DMA (DMA)  methods for TX and RX operations ranging from 1 to 100 bytes length,
    or 1 to 100 samples for ADC. Figure shows a relatively high overhead for small transfer size and low or negative overhead 
    for larger transfers. }
    \label{fig:overhead}
\end{figure*}

To understand the overall effects of \sysname on peripheral operations, we
measured and compared the overhead and the core processor usage when a task \textbf{read}
(RX) or \textbf{write} (TX) data streams on peripherals using pooling, interrupts, and
insecure DMA methods (i.e., we looked at the three possible methods used by
firmware to communicate with peripherals).

\textbf{Setup:} for pooling, interrupt and insecure DMA methods, we configured
\fmpu with user-defined regions granting access to the tested peripheral and the
\dctr. For \sysname, we configured it with a capability granting read or write
privileges on the tested peripheral.  We simulated diverse conditions by reading
or writing data streams ranging from 1 to 100 bytes on USART, I2C, and SPI
peripherals. In the case of ADC, we performed conversions ranging from 1 to 100
samples on different channels.  For each data length/sample size, we collected
the overhead 10 times and averaged the result. For I2C and SPI, we used an
extra development board connected as a dummy slave to our tested device.  For
USART, we used our workstation connected to the development board through the
ST-Link USART bridge.

We measured the overhead by counting the number of cycles using similar
instrumentation as described in \S\ref{sec:MicroBenchmark}.  To measure the core
processor usage, we used more sophisticated instrumentation provided by Percepio
Tracealyzer \cite{percepioPercepioTracealyzerStop}. 
We also added a secondary communication channel between the MCU and our
workstation through a USART and a USB-to-USART bridge. We used this channel for
synchronization and collection of our experiments' results.

\textbf{Results:} the performance overhead of \sysname is inversely proportional
to the data length. In Figure \ref{fig:overhead}, we observe that the overhead
can be as high as 134\% when transmitting a single byte through SPI (SPI TX),
and, in the same operation, the overhead is below 10\% and trending towards zero
when transmitting 50 or more bytes.  The reason for this behavior is the costly
initialization of a DMA transfer.  This initialization includes the policy
verification and the \dctr configuration per transfer (i.e., that the overhead
is constant and added per transaction, and not per byte).  In the case of ADC,
\sysname presents a small positive overhead (4.71\% on average) only when
compared to the insecure DMA method. The reason for this behavior is the
highly-DMA-optimized ADC peripheral populated on our tested MCU. Without DMA,
the ADC operations suffer a considerable overhead for IT and Pool methods. The
USART RX operation is also a particular case because our workstation's jitter
disguises the overhead of \sysname.  The workstation's timing is not comparable
to a real-time device such as the used MCU. We include complete details of the
maximum, average and minimum performance results in appendix \S\ref{ap:overhead}.

\sysname maintains the scheduler timing constraints adding only 2\% of core
processor usage compared to insecure DMA on average. Also, it reduces core
processor usage by 35.9\% and 33.7\% compared to pooling and interrupt
methods, respectively (Table \ref{table:cpu_usage}).  The considerable reduction
of core usage for both DMA methods is due to the \dctr taking care of the data
movements between peripheral and RAM on behalf of the core processor.

\begin{table}[]
  \scriptsize 
  \centering
  \begin{tabular}{l|l|ll|ll|ll}
  \multirow{2}{*}{\textbf{Operation}} & \textbf{D-BOX} & \multicolumn{2}{c|}{\textbf{Pool}}     & \multicolumn{2}{c|}{\textbf{IT}}       & \multicolumn{2}{c}{\textbf{DMA}}       \\ \cline{2-8} 
                                      & \textbf{[\%]}  & \textbf{[\%]} & \textbf{$\Delta$ } & \textbf{[\%]} & \textbf{$\Delta$ } & \textbf{[\%]} & \textbf{$\Delta$ } \\ \hline
  \textbf{USART TX}                   & 13.5           & 68.5          & -55.0                  & 32.6          & -19.1                  & 12.4          & 1.1                    \\
  \textbf{USART RX}                   & 13.3           & 80.6          & -67.3                  & 13.5          & -0.2                   & 10.9          & 2.4                    \\
  \textbf{SPI TX}                     & 18.6           & 45.6          & -27.0                  & 49.8          & -31.2                  & 14.0          & 4.6                    \\
  \textbf{SPI RX}                     & 18.1           & 45.5          & -27.4                  & 65.2          & -47.0                  & 14.0          & 4.2                    \\
  \textbf{I2C TX}                     & 17.5           & 40.5          & -22.9                  & 52.8          & -35.3                  & 15.8          & 1.7                    \\
  \textbf{I2C RX}                     & 17.8           & 40.1          & -22.3                  & 54.7          & -37.0                  & 21.6          & -3.8                   \\
  \textbf{ADC}                        & 20.3           & 49.9          & -29.6                  & 86.1          & -65.8                  & 16.6          & 3.7                    \\ \hline
  \textbf{Average}                    & 17.0           & 52.9          & -35.9                  & 50.7          & -33.7                  & 15.0          & 2.0                   
  \end{tabular}
  \caption{Core processor usage [\%] for peripheral operations relative to the total core capacity. 
  Lower values mean better management of tasks and guarantee of timing constraints. $\Delta$ is the processor usage difference compared with \sysname. }
  \label{table:cpu_usage}
\end{table}

\subsection{Power analysis}
\label{sec:powerenergy}

\textbf{Setup:} For this analysis, we used the same software setup of
\ref{sec:PeripheralPerformance}, but without the instrumentation. We only kept
the secondary USART channel for synchronization purposes.  For each test, we
modified the firmware to configure and provide power exclusively to the tested
peripheral and the secondary USART channel, keeping the rest of the tested
peripherals unpowered.  We executed the Read/Write operations for USART, I2C,
SPI, and ADC, ranging from 1 to 100 bytes/samples in a continuous loop while
taking samples of current and voltage every 100 ms, totaling 1000 samples.  To
measure the current and voltage, we used the highly accurate power monitor
INA226 \cite{texasinstrumentsINA226HighSideLowSide} connected to our workstation
through the USB-to-I2C bridge MCP2221A \cite{microchipMCP2221AUSBUSART}. We
configured the electrical connections of the INA226 and the development board to
measure exclusively the current and voltage applied to the STM32L152RE MCU.

\begin{table}[]
\scriptsize 
\centering
\begin{tabular}{l|l|ll|ll|ll}
  \multirow{2}{*}{\textbf{Operation}} & \textbf{D-BOX}      & \multicolumn{2}{c|}{\textbf{Pool}}                 & \multicolumn{2}{c|}{\textbf{IT}}                   & \multicolumn{2}{c}{\textbf{DMA}}                   \\ \cline{2-8} 
                                      & { \textbf{[mW]}} & { \textbf{[mW]}} & { \textbf{$\Delta$ [\%]}} & { \textbf{[mW]}} & { \textbf{$\Delta$ [\%]}} & { \textbf{[mW]}} & { \textbf{$\Delta$ [\%]}} \\ \hline
  \textbf{USART TX}                   & 31.0                & 37.9                & -18.3                        & 38.1                & -18.8                        & 38.7                & -19.9                        \\
  \textbf{USART RX}                   & 30.4                & 37.2                & -18.3                        & 37.6                & -19.1                        & 38.8                & -21.6                        \\
  \textbf{SPI TX}                     & 31.1                & 35.8                & -13.2                        & 37.1                & -16.3                        & 37.5                & -17.1                        \\
  \textbf{SPI RX}                     & 31.0                & 36.9                & -15.9                        & 37.1                & -16.6                        & 36.3                & -14.8                        \\
  \textbf{I2C TX}                     & 31.5                & 37.9                & -16.8                        & 38.1                & -17.3                        & 38.1                & -17.3                        \\
  \textbf{I2C RX}                     & 31.6                & 37.7                & -16.2                        & 38.1                & -17.0                        & 38.3                & -17.4                        \\
  \textbf{ADC}                        & 31.8                & 36.2                & -12.0                        & 38.6                & -17.4                        & 39.4                & -19.2                        \\ \hline
  \textbf{Average}                    & 31.2                & 37.1                & -15.8                        & 37.8                & -17.5                        & 38.1                & -18.2                       
  \end{tabular}
  \caption{Average power [mW] required for peripheral operations. $\Delta$ [\%] is the relative power usage difference compared with \sysname. }
  \label{table:power}
\end{table}

\textbf{Results:} \sysname reduces the power requirements on average 15.8\%,
17.5\%, and 18.2\% compared to pooling, interrupt, and insecure DMA methods,
respectively (Table \ref{table:power}).  The result of insecure DMA seems
counter-intuitive at first glance since it is expected to reduce power
consumption compared to IT and Pool methods.  However, we determined that the
vendor's DMA HAL (Hardware Abstraction Layer) used for insecure DMA is not power
efficient. On the other hand, \sysname implementation uses the DMA driver
directly and replaces the HAL with the DMA task of our design.  Our lean
implementation demonstrates that DMA can be secure, and significantly reduce
power consumption---a highly desirable characteristic for embedded and
IoT applications. We include a graphic representation of our results in appendix 
\S\ref{ap:energy}.

\subsection{Memory overhead analysis}
\label{sec:memoryanalysis}

\textbf{Setup:} To analyze the \sysname memory overhead, we compared the
firmware images of \fmpu and \sysname used in \ref{sec:powerenergy}. We also
break down the specific requirements of our prototype implementation.  We
compiled all images with no optimization (-O0) using the GNU Tools (release
7-2018-q2) included in the CubeIDE version 1.5.0.

\textbf{Result:} \sysname, on average, reduces the usage of flash and RAM by
-0.12\% and -0.07\%, respectively (Table \ref{table:memoryoverhead}). This
reduction is due to differences in vendors' driver libraries. \fmpu uses a
feature rich HAL library, whereas \sysname uses ``low-level'' drivers. Note that
the RAM requirements are constant for both systems. This is because FreeRTOS
reserves and manages its heap statically (i.e., \sysname's RAM requirements are
part of the already reserved heap observable by the compiler).  Besides this
FreeRTOS characteristic, Table \ref{table:memorysys} details mandatory and
driver-independent memory requirements of our \sysname prototype implementation.
Notice that \sysname RAM and flash requirements depend on the number of channels
supported by the \dctr and the number of policy's capabilities, respectively. In
our prototype, the security policy supports 3 capabilities that require 36 bytes
in total (12 bytes per capability). In terms of RAM, the DMA request queue needs
320 bytes (32 byte per DMA channel).

    \begin{table}[] \scriptsize \centering \begin{tabular}{l|lll|lll} &
\multicolumn{3}{c|}{\textbf{RAM [kB]}}               &
\multicolumn{3}{c}{\textbf{Flash [kB]}}              \\ \hline
\textbf{Peripheral} & \textbf{\fmpu} & \textbf{\sysname} & \textbf{$\Delta$
[\%]} & \textbf{\fmpu} & \textbf{\sysname} & \textbf{$\Delta$} [\%]\\ \hline
\textbf{USART}      & 40.61          & 40.58             & -0.07           &
129.68         & 129.7             & 0.02            \\ \textbf{I2C}        &
40.61          & 40.58             & -0.07           & 128.66         & 128.98 &
0.25            \\ \textbf{SPI}        & 40.61          & 40.58             &
-0.07           & 128.34         & 128.32            & -0.02           \\
\textbf{ADC}        & 40.61          & 40.58             & -0.07           &
129.45         & 128.05            & -1.08           \\ \hline \textbf{Average}
& 40.61          & 40.58             & -0.07           & 129.03         & 128.88
& -0.12           \\   \end{tabular} \caption{\sysname RAM and flash overhead
compared to \fmpu.} \label{table:memoryoverhead} \end{table}
    
\begin{table}[] 
\scriptsize 
\centering 
\begin{tabular}{l|ll} & \textbf{RAM
{[}Bytes{]}} & \textbf{Flash {[}Bytes{]}} \\ \hline 
\textbf{Privileged Data}           &230                       & --                          \\ 
\textbf{DMA request queue}   & 320 & --                          \\
\textbf{System calls}              & --                       & 916                         \\
\textbf{DMA task}                  & 1024                     & 1648                        \\
\textbf{TCB capabilities} & --                        & 36     \\ \hline 
\textbf{Total}                     & 1574                     & 2600                      
\end{tabular} \caption{\sysname RAM and flash
requirements } \label{table:memorysys} \end{table}

\subsection{Case study: Securing a Programmable Logic Controller }
\label{sec:caseStudies} In this section, we present a representative case study
that demonstrates how \sysname can improve the security, performance and power
consumption in a real-world scenario.

\subsubsection{Programmable Logic Controllers}

PLCs are embedded, ruggedized computers used in industrial environments to
control critical processes.  These devices operate continuously, assuring
availability, reliability and performance. Due to PLC's relationship with
critical infrastructure, these devices have been the targets of numerous
cybersecurity attacks; usually, resulting in devastating damages
\cite{dipintoTRITONFirstICS2018,falliereW32StuxnetDossier2010,sansinstituteConfirmationCoordinatedAttack2016,cisaHavexICSAlert2014}.

The PLC analyzed in this case study is the Wecom LX3VM 2424M \cite{WeconLX3VM2424M}. 
This PLC has been used in water supply facilities \cite{WeconSmartWater}, 
poultry processing plants \cite{WeconAutomaticSolution}, 
refrigeration applications \cite{WeconAutomaticControl}, and small
machines as we describe next.

\subsubsection{Firmware characteristics}

The firmware used in this section
contains open source libraries and a subset of proprietary code that we
acquired from an industrial connection.  We adapted the code to run on our
development board instead of the original PLC
to facilitate metrics evaluation. Nevertheless, 
the PLC and our development board integrate similar
MCUs from ST Microelectronics.

\begin{figure}
        \centering
        \includegraphics[width=6cm]{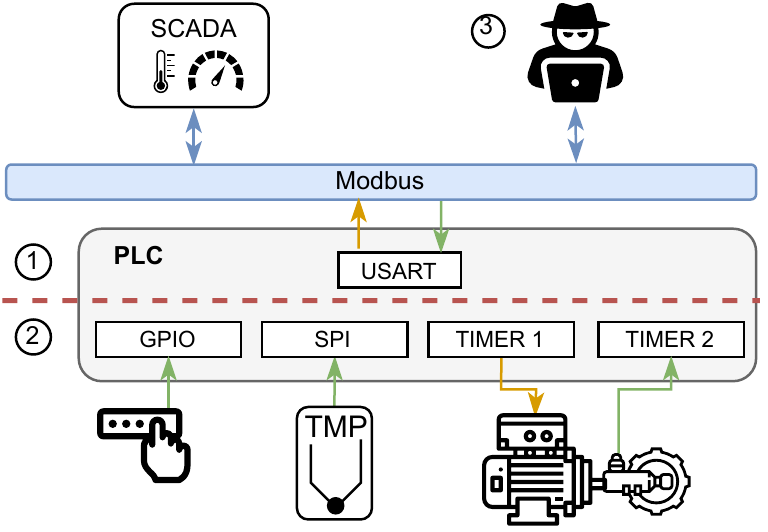}
        \caption{Compartmentalization of a PLC-Based motor control system. \Circled{1} Modbus protocol for SCADA monitoring, 
        \Circled{2} Push buttons, thermocouple, motor drive and encoder connected to on-chip peripherals, 
        \Circled{3} Attacker abusing the Modbus protocol.}
        \label{fig:modbuscase}
    \end{figure}

The PLC firmware uses FreeRTOS to implement a motor control system for an
automatic molding injection machine---this is an in-house development 
that replaced the original firmware provided by the PLC vendor. 
This firmware scans inputs from a rotary encoder (speed feedback), 
thermocouple conditioner (temperature), and push
buttons (local activation). It then executes a PID
(Proportional-Integral-Derivative) control loop, and updates the output using
PWM (Pulse-Width Modulation) to activate the motor drive (\Circled{2} in Figure
\ref{fig:modbuscase}).  The firmware also integrates the Modbus RTU protocol for
remote communication with an SCADA (Supervisory Control And Data Acquisition)
system (\Circled{1} in Figure \ref{fig:modbuscase}).  The PLC has a 10 ms scan
cycle (i.e., scanning inputs, executing control logic, and updating outputs must
occur every 10 ms).

\subsubsection{Threat analysis}

The original firmware is insecure because it does not enforce the least
privilege principle. The tasks have access to all data, code and peripherals.
There is no distinction between kernel and user space. A bug or a backdoor in
the Modbus protocol, for example, can be exploited remotely by an attacker
(\Circled{3} in Figure \ref{fig:modbuscase}) to hijack the PLC operation and
perform malicious actions, causing destruction of facilities, economic loss, or
even personal injury.

\subsubsection{Solution} We compartmentalized the PLC firmware by porting it to
\sysname using two unprivileged tasks, and with minimal engineering effort.
First, we modified the Modbus protocol to run in an unprivileged task using two
user-defined regions and a single DMA capability granting read and write rights
on USART. The user-defined regions contain the Modbus protocol handlers, and a
shared region between the PLC task and the Modbus task.  Second, we modified the
PLC scan cycle to run in an unprivileged task using three user-defined regions,
a DMA capability and an ISR. The first user-defined region is the shared region
with the Modbus task. In this region, the PLC task safely exposes data to the
Modbus protocol. The second and the third user-defined regions grant access to
the Timer 1 and GPIO peripherals, respectively.  
The DMA capability with EAR grants read access to the off-chip thermocouple
conditioner connected through SPI using DMA.

\subsubsection{Evaluation and analysis}

We tested the PLC firmware in our development board connected to our
workstation, simulating the SCADA system with a Modbus client.  The
SCADA system sets the PID control loop parameters and the motor speed
on the PLC (development board). Then, the SCADA system continuously
monitors the speed and temperature of the motor through the Modbus
protocol. We collected security, performance, and energy metrics using
similar procedures as described in \S\ref{sec:PerformancelAnalysis} and
\S\ref{sec:powerenergy}.

\sysname improves the PLC security properties, maintains the real-time
constraint, and reduces the power and CPU usage as detailed in Table
\ref{table:casemodbus}. 
Remote attacks exploiting vulnerabilities in Modbus are contained and will only
affect the Modbus task.  The compartmentalization will keep the critical PLC task working
even when an attacker takes control of the Modbus task and misuses its DMA
capability.  This case study demonstrates that by securely using the DMA
controller, is possible to protect MCU resources while maintaining or improving
the performance and power requirements of a practical embedded application.

\begin{table}[]
    \scriptsize
        \centering
    \begin{tabular}{l|lll}
      & \textbf{FreeRTOS} & \textbf{\sysname} & \textbf{$\Delta$ } \\ \hline
    \textbf{Flash ratio [\%]}      & 100               & 52                & -48.00              \\
    \textbf{RAM ratio [\%]}       & 100               & 25.20              & -74.80             \\
    \textbf{Peripheral ratio [\%]} & 100               & 0.32              & -99.68             \\ 
    \textbf{ROP gadgets [\#]}     & 0                 & 0                 & 0                 \\ \hline
    \textbf{Avg. Scan cycle [mS]}   & 9.99             & 10                & 0.01               \\
    \textbf{Avg. power [mW]}      & 41.38             & 34.94             & -15.56              \\  
    \textbf{CPU usage [\%]}       & 16.79             & 12.96            & -3.83

    \end{tabular}
    \caption{Evaluation of PLC firmware implementing \sysname compartmentalization. Lower ratio and ROP represents a better protection.
    $\Delta$ calculated as absolute difference for percentages, and as relative difference for other units.}
    \label{table:casemodbus}
    \end{table}

\section{Discussion}
\label{sec:discussion}

\subsection{The capability-based security model properties and its limitations}

The capability-based security model assures two properties: 1) the deputy (DMA
task) has no rights to the resources that are affected by a DMA transfer, and 2)
it guarantees that a task cannot request DMA operations on a resource without
proper rights. 

Since the resource and its rights are inseparable, immutable and owned by an
identifiable task, it is possible to avoid confused deputy problems that can
affect DMA operations by design. However, it is still possible to write insecure
applications canceling any of the properties of the capability-based model. For
example, if two mutually independent functionalities with different security
profiles are structured as a single task, there is no way for the DMA task to
discriminate the ``correct'' rights on resources. Clearly, this example is an
obvious design error. At the same time, it is common to find these insecure
implementation patterns affecting more than just the DMA operations.

\subsection{\sysname trusted computing base}
\sysname TCB includes the kernel, the DMA task (drivers and 
addressing routines) running at the privileged level. This 
characteristic requires special care because of the associated 
risk with a relatively large code base. 

\sysname's separation of the DMA task and kernel supports running the DMA task 
in unprivileged mode by design, thus mitigating its associated risk. 
However, running the drivers in unprivileged mode, with 
only three MPU user-defined regions, requires considerable engineering 
effort and would make the solution vendor-specific. Regardless of the 
engineering effort, developers should consider running the DMA task in 
unprivileged level when the risk profile of the embedded application 
requires the highest protection.

\subsection{The ARMv8-M security extensions} The ARMv8-M includes hardware
support to mitigate many of the issues related to multimaster environments,
including DMA. Notably, this architecture provides a bus-level security feature
that can verify and propagate the permissions for each bus operation. In this
case, the properties of memory regions are enforced system-wide and not
exclusively on a specific master interface \cite{HighendSecurityFeatures2017}.
In addition, ARMv8-M provides the newly added "Test Target" (TT) instruction. 
This instruction accelerates the verification of security access permissions 
of memory ranges, which is essential to validate DMA operations.

The new system-level security accompanied by TrustZone and a more flexible MPU
of the ARMv8-M bring tremendous security improvements as well as performance for
low-end devices. Especially when compared to software-based solutions like \sysname.
Nevertheless, the design considerations of \sysname are still
helpful to implement secure firmware architectures that take advantage of the
new hardware security features.

\subsection{Compatibility of \sysname with other solutions} \sysname procedures
can be applied to other compartmentalization solutions that define the same
compartmentalization unit (granularity) and implement RTOS designs. For example, MINION
\cite{kimSecuringRealTimeMicrocontroller2018} defines its memory view at a
thread level, and it is built on top of NuttX RTOS.  On the other hand,
solutions such as ACES \cite{clementsACESAutomaticCompartments2018} are not
directly compatible (i.e., ACES defines compartments at a function level for
firmware implementing baremetal designs). This last design makes it very
challenging to validate sources and destinations for DMA operations because the
compartment views are the result of an algorithm that does not capture the
semantics of operations occurring on a different master (i.e., the \dctr).

\section{Related Work}

Our work intersects DMA protection and resource compartmentalization in embedded
devices.

\subsection{DMA protection}

Existing works ignore or are limited in supporting secure DMA transfers in embedded devices. 
EPOXY \cite{clementsProtectingBareMetalEmbedded2017} does not consider DMA as a security-sensitive operation, and 
its method automatically elevates privileges when accessing a peripheral like the \dctr.
TockOS \cite{levyMultiprogramming64kBComputer2017} and uVisor \cite{armARMMBEDUvisor2021} acknowledge the 
security-critical behavior of the DMA controller, 
but provide only partial support for secure DMA transfers. 
TockOS uses the MPU to enforce memory access, and leverages Rust safe types  to protect
the DMA controller. 
However, TockOS does not consider the DMA descriptors stored in RAM, which conversely, \sysname protects as security critical structures. 
uVisor \cite{armARMMBEDUvisor2021} recommends enforced access to the DMA controller through SVC-based APIs, 
but this reference design is deprecated. 
Trusted Firmware-M \cite{trustedfirmware.orgTrustedFirmware} works in a similar fashion and targets exclusively TrustZone-enabled devices (ArmV8-M)
or dual-core enabled devices. \sysname targets ArmV7-M devices and considers multimaster environments.

To protect DMA as a general master I/O device, 
Haglund et. al presented a mechanism \cite{haglundTrustworthyIsolationDMA2019} based on a runtime monitor to isolate I/O devices from accessing sensitive memory regions. 
Haglund's work is limited to NIC operations over DMA. 
In the context of general purpose computers, several works \cite{willmannProtectionStrategiesDirect2008,zhouBuildingVerifiableTrusted2012,zhouOndemandIsolatedSecuritysensitive2014,yuSeparationModelIts2019}
 presented security policies and protection strategies on secure access of master I/O devices. 
Compared to \sysname, these works do not consider the hardware availability and constraints of embedded
devices.  Prior work has also considered DMA implications in the security of remote 
attestation \cite{nunesVRASEDVerifiedHardware2019}, the integrity of software 
execution \cite{nunesAPEXVerifiedArchitecture2020}, and undetectable 
firmware modifications through DMA abuse \cite{nunesTOCTOUProblemRemote2020}. 
These works require small hardware modifications, but, similarly to D-Box, 
they expose the need to systematically protect DMA to guarantee the operation and security
of deeply embedded devices.

\subsection{Compartmentalization}
\subsubsection{Software-based memory compartmentalization}
Existing works \cite{clementsProtectingBareMetalEmbedded2017,clementsACESAutomaticCompartments2018,hardinApplicationMemoryIsolation2018}
leverage embedded compilers to create two or more isolated execution environments. 
EPOXY \cite{clementsProtectingBareMetalEmbedded2017} 
overlays the MPU to enforce different permissions and execution levels. %
It creates two distinct domains based on the execution levels.  
However, EPOXY requires manual annotations and the execution levels cannot be statically identified. 
ACES \cite{clementsACESAutomaticCompartments2018} creates an instrumented binary to enforce runtime compartmentalization.  
Similar to \sysname, ACES also takes a developer-specified policy. 
However, ACES requires a complex graph traversal algorithm for policy definition.
In contrast, Hardin et al. \cite{hardinApplicationMemoryIsolation2018} leveraged compiler-inserted code to achieve runtime bound-checking. %
This technique is based on AmuletOS
\cite{hesterAmuletEnergyEfficientMultiApplication2016}, %
and it can reduce runtime overhead in dynamic checking.  
However, it still depends on language features and clear OS rules. 
NesCheck \cite{midiMemorySafetyEmbedded2017} 
provides compartmentalization %
by extending the existing TinyOS compiler toolchain~\cite{deckerTinyOS} with LLVM-based passes, 
and focuses on runtime checks on nesC programs. %
Compared with \sysname, compiler-based compartmentalization may involve
imprecise analysis to form compartments and cumbersome policy definitions. 
\sysname makes it easy to define static policies by developers. 

Without leveraging compilers, 
TockOS \cite{levyMultiprogramming64kBComputer2017} 
divides the kernel into a trusted core for critical tasks, and a non-trusted capsule for
peripheral drivers and non-system-critical tasks.  
MINION \cite{kimSecuringRealTimeMicrocontroller2018} constructs a
per-process memory view by an offline clustering analysis on the
system firmware.  
MINION is similar to \sysname in using a thread/task as the unit of compartmentalization. 
However, MINION is inaccurate in approximating the resources required for each thread. 
MINION's inaccuracy can lead to unexpected crashes in production environments, whereas \sysname
explicit resource definitions assure application functionality and security.

In addition to MPU-assisted compartmentalization, the uXOM
\cite{kwonUXOMEfficientEXecuteOnly2019} uses unprivileged memory
instructions to enable efficient execute-only-memory protections on ARM Cortex-M devices.
Mate~\cite{levisMateTinyVirtual2002} isolates applications through virtualizing the single memory space. 
SANCTUARY \cite{brasserSANCTUARYARMingTrustZone2019} leverages ARM
TrustZone Address-Space Controller
to enforce hardware-level isolation.  
These solutions rely on specific hardware, or provide orthogonal protections
that can be integrated in a compartmentalization solution where
\sysname is implemented. 
In the context of general purpose computers, 
existing works \cite{chenShredsFineGrainedExecution2016,hsuEnforcingLeastPrivilege2016} provide programming primitives to divide and control accesses to the memory.
Shreds \cite{chenShredsFineGrainedExecution2016} 
isolates fine-grained program segments %
from others in the same process.  
It uses the compiler toolchain and the OS module to enable the segment execution.  
SMV \cite{hsuEnforcingLeastPrivilege2016} aims at per-thread access control, 
and divides the virtual memory into multiple domains, and enforces the least privilege principle. 
Compared to \sysname, 
these solutions will need to resolve the performance and hardware constraints
before they can be applied to 
embedded devices.

\subsubsection{Hardware-assisted memory isolation}
Existing works extend hardware to improve memory isolation.
Strackx et. al \cite{strackxEfficientIsolationTrusted2010} presented
self-protecting modules (SPM) to isolate trusted subsystems sharing
the same processor and memory space.  They created three hardware
instructions %
to manage the SPM, and provided a memory access control model for subsystems using the program counter. 
However, SPM must exclude protected memory locations that are accessed by DMA.
Toubkal \cite{sensaouiToubkalFlexibleEfficient2019} 
enhances the MPU by adding a new hardware layer to create different access environments for different hardware components.  
Trustlite \cite{koeberlTrustLiteSecurityArchitecture2014} presents an execution-aware MPU (EA-MPU) to execute trusted modules.  
It extends the MPU by providing a means to link code regions to data regions, 
and validate the address of the executing instruction.  
Trustlite can isolate data of each module from the other parts of the program.  
TyTan \cite{brasserTyTANTinyTrust2015} also protects the memory with an EA-MPU. 
It isolates tasks with a secure IPC proxy task. 
Hardware-assisted implementations allow more flexible MPU configuration and better interrupt handling. 
However, these solutions usually target the generic memory compartmentalization problem,  
without considering developer-supplied policies nor the least privilege principle. 
The changes to the hardware are also not available on commercial devices. In
contrast, 
\sysname only uses standard and broadly available hardware.

In the context of x86 CPUs, existing works
\cite{zhangSoKStudyUsing2016} leveraged various mechanisms for memory
compartmentalization and protection, including MMU, MPK, Intel management engine,
and Intel SGX. For example, IMIX \cite{frassettoIMIXInprocessMemory2018} provides a lightweight
in-process memory compartmentalization.  It extends the x86 ISA with a security-sensitive memory-access permission.
Unlike \sysname, these solutions are not available on low-end hardware of embedded devices.

\section{Conclusion} \label{sec:conclusion}

In this paper, we presented \sysname, a systematic approach to enable secure DMA
operations for compartmentalization solutions of embedded applications---a
problem that has been largely ignored that we analyze and expose.  \sysname
defines a reference architecture and a workflow to holistically protect the DMA
life-cycle.  \sysname uses a capatibility-based security model to provide strong
protections, and pragmatic methods to define a DMA policy compatible with MCU
software and hardware constraints.

To evaluate a prototype of \sysname implemented on top of \fmpu, we performed
qualitative and quantitative analyses of different benchmark programs.  Our
results show that \sysname provides secure DMA access while reducing the attack
surface of \fmpu for all the 6 security metrics that we used. \sysname incurred
a low overhead for kernel and peripheral operations while reducing the overall
power requirements. By testing on a real-world PLC application, we further
confirmed \sysname's security improvement, low overhead, and reduced power
consumption thanks to its support for secure DMA transfers, and a lean
implementation.  Lastly, we discuss \sysname's limitations and its compatibility
with existing security solutions.

\section*{Acknowledgment}

The authors would like to thank the anonymous 
reviewers and our shepherd Ivan De Oliveira Nunes 
for their insightful comments. 
This project was supported
by the National Science Foundation (Grants\#: CNS-1748334, and CNS-2031390),
and the Office of Naval Research (Grant\#: N00014-18-1-2043).
Any opinions, findings, and conclusions or recommendations
expressed in this paper are those of the authors and do not
necessarily reflect the views of the funding agencies.

\bibliographystyle{plain}
\bibliography{../../bibtex/EmbeddedCompartments}
\appendix
\label{ap:appendix}
\subsection{Survey on availability of hardware security features on MCUs}
\label{sec:survey}

The main isolation components of the Cortex-M embedded architecture
are the MPU and TrustZone. These two
components are optional, and their support depends on the MCU vendor
implementation. Also, each component is only supported by specific
Cortex-M micro-architecture profiles. The MPU is supported by Armv6-M
(Cortex-M0+), Armv7-M (Cortex-M3/M4/M7) and Armv8-M
(Cortex-M23/M33/M55), whereas TrustZone is exclusively supported by
the Armv8-M (Cortex-M23/M33/M55).

\begin{figure}
    \centering
    \includegraphics[width=8.5cm]{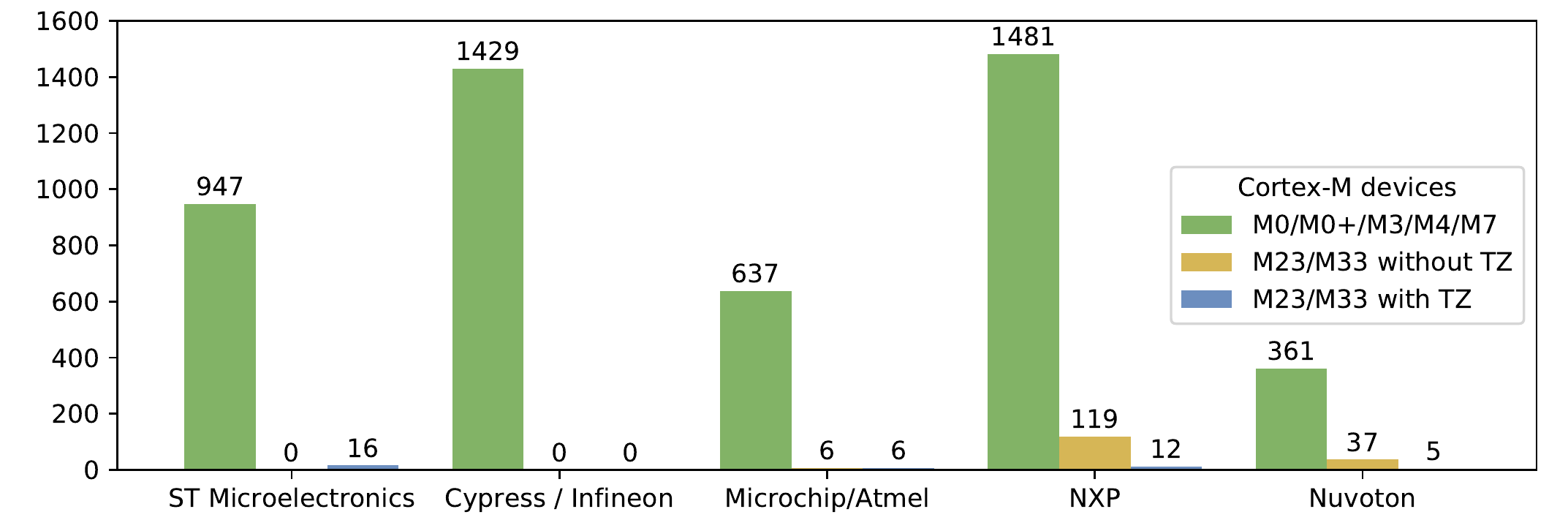}
    \caption{ ARM TrustZone (TZ) availability in the portfolio of five top global MCU Vendors in February 2021 .
    \cite{cypressMicrocontrollerPortfolioInfineon,microchipNewPopular32bit,NuvotonM261M262,nxpProductSelectorNXP,stmicroelectronicsSTMCUFINDERPCSTM32STM8}.  }
    \label{fig:TZavailablity}

\end{figure}

To verify the current availability of the ARM TrustZone feature in
commercial MCU devices, we surveyed the complete portfolio of the top
five global MCU vendors. The results depicted in Figure
\ref{fig:TZavailablity} demonstrates that TrustZone-enabled devices
are still scarce. In the best case, ST Microelectronics commercializes
16 parts which represents only 1.6\% of its portfolio. In contrast,
Cypress/Infineon has no Armv8-M commercial devices at all. Also, no
vendor currently offers commercial devices for the Cortex-M55
micro-architecture. The current availability is slightly better compared
to the total absence of devices supporting this technology around
2017-2018 as observed by \cite{clementsACESAutomaticCompartments2018}.

\begin{figure}
    \centering
    \includegraphics[width=5cm]{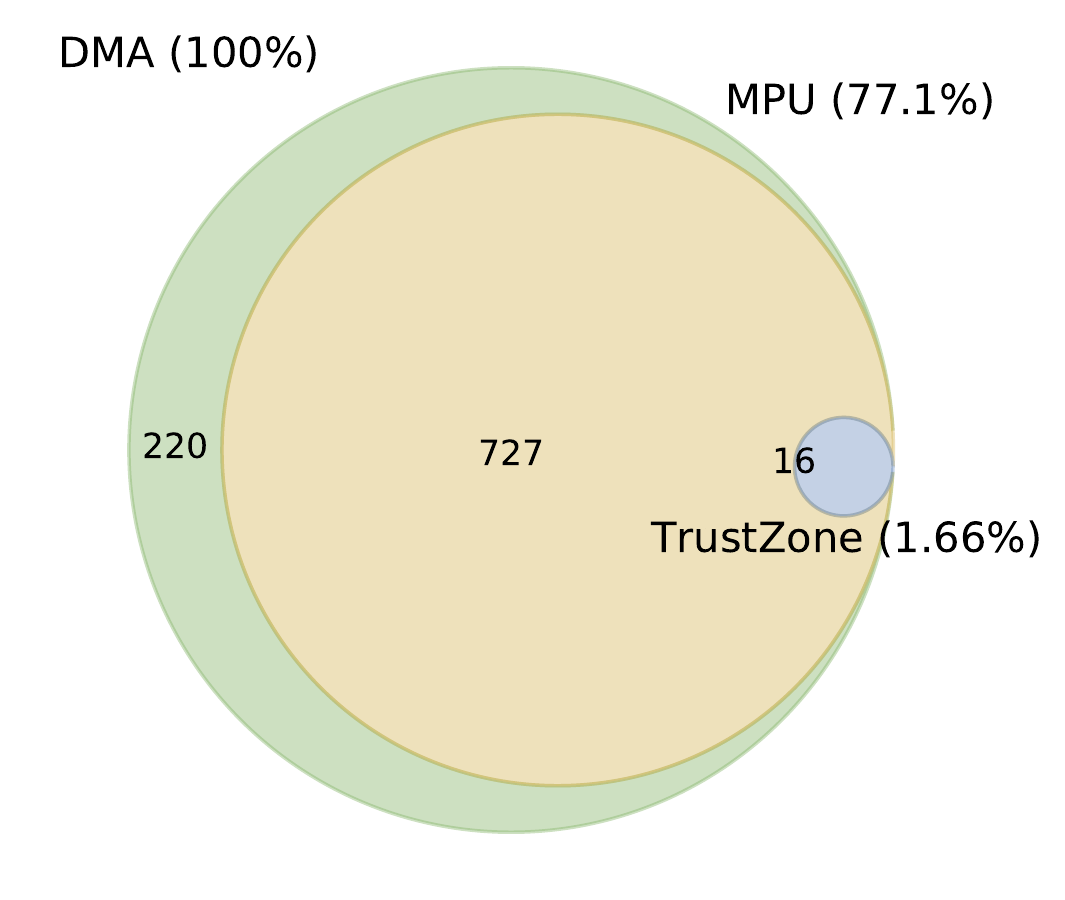}
    \caption{ MPU and \dctr availability in the ST Microelectronics portfolio in February 2021.  \cite{STMCUFINDERSTM32STM8}.  }
    \label{fig:DmaMPUtz}

\end{figure}

\subsection{Power requirements for peripheral operations}
\label{ap:energy}
\begin{figure*}[t!]
    \centering
    \includegraphics[width=18cm]{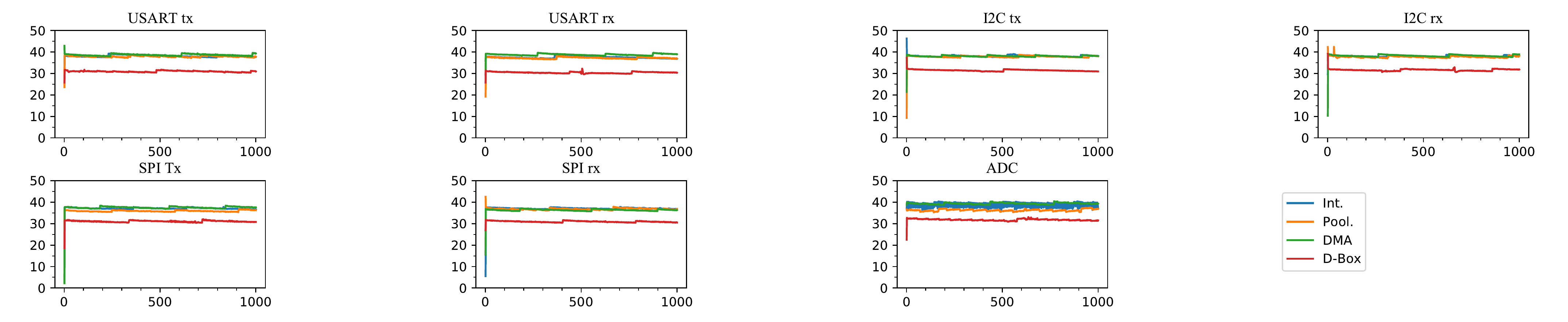}
    \caption{Power requirements of peripheral operations observed over 1000 samples using Pooling, Interrupts, Insecure DMA, and \sysname methods. }
    \label{fig:energy}
\end{figure*}

Figure \ref{fig:energy} depicts the power consumption recorded in a synchronized continuous loop. 
In all the cases DMA operations using \sysname procedures are more efficient.

\makeatletter
\setlength{\@fptop}{0pt}
\makeatother

\label{ap:PeripheralPerformance}

\subsection{Performance overhead details}
\label{ap:overhead}
Table \ref{table:performance} details the minimum, average, and maximum performance
overhead observed in our experiments with USART, I2C, SPI, and ADC peripherals.
In general, the use of DMA is optimized for larger transfers. We observed in all cases
that the overhead is less than 10\% when the transfer is 25 or more bytes long.
Transfers involving more than 25 bytes are common for communication protocols (e.g., 
the MQTT protocol, which is widely used by IoT devices).

\begin{table}[]
    \begin{tabular}{l|lll|lll|lll}
    \multirow{2}{*}{\textbf{Operation}} & \multicolumn{3}{c|}{\textbf{Pool [\%]}}    & \multicolumn{3}{c|}{\textbf{IT [\%]}}      & \multicolumn{3}{c}{\textbf{DMA [\%]}}      \\ \cline{2-10} 
                                        & \textbf{Min} & \textbf{Avg} & \textbf{Max} & \textbf{Min} & \textbf{Avg} & \textbf{Max} & \textbf{Min} & \textbf{Avg} & \textbf{Max} \\ \hline
    \textbf{USART-TX}                   & 0.8          & 3.8          & 46.3         & 0.3          & 1.5          & 16.0         & 0.3          & 1.2          & 13.3         \\
    \textbf{USART-RX}                   & -2.4         & -0.2         & 2.9          & -2.9         & -0.6         & 1.6          & -2.7         & -0.1         & 2.8          \\
    \textbf{SPI-TX}                     & 2.6          & 10.4         & 134.5        & 1.0          & 4.6          & 73.2         & 0.9          & 4.3          & 72.8         \\
    \textbf{SPI-RX}                     & -7.6         & 0.3          & 114.0        & 1.3          & 4.7          & 54.2         & 0.8          & 3.0          & 35.5         \\
    \textbf{I2C-TX}                     & -10.0        & -1.1         & 103.3        & -0.5         & 1.7          & 38.6         & 10.9         & 12.3         & 45.6         \\
    \textbf{I2C-RX}                     & 1.9          & 9.4          & 89.7         & -0.7         & 1.6          & 33.9         & 0.6          & 8.0          & 14.5         \\
    \textbf{ADC}                        & -80.4        & -67.9        & 88.4         & -86.9        & -77.2        & 72.7         & 1.1          & 4.2          & 34.0         \\ \hline
    \textbf{Average}                    & -13.60       & -6.48        & 82.73        & -12.63       & -9.12        & 41.46        & 1.70         & 4.71         & 31.21       
    \end{tabular}
    \caption{Performance overhead of \sysname versus Pooling (Pool), Interrupt (IT) and Insecure DMA (DMA) methods.}
    \label{table:performance}
    \end{table}

\subsection{Acronyms}
\label{ap:acronyms}

\textbf{ADC} - Analog to Digital Converter.

\textbf{CAN} - Controller Area Network.

\textbf{GPIO} - General Purpose Input/Output.

\textbf{HAL} - Hardware abstraction layer.

\textbf{I2C} - Inter-Integrated Circuit.

\textbf{MMIO} - Memory-mapped Input/Output.

\textbf{MQTT} - Message Queuing Telemetry Transport.

\textbf{PLC} - Programmable Logic Controller.

\textbf{ROP} - Return Oriented Programming.

\textbf{SCADA} - Supervisory Control And Data Acquisition.

\textbf{SPI} - Serial Peripheral Interface.

\textbf{SVC} - Supervisor Call.

\textbf{TCB} - Thread Control Block.

\textbf{USART} - Universal Synchronous/Asynchronous Receiver/Transmitter.

\end{document}